\documentclass[11pt,a4paper]{article}
\pdfoutput=1
\usepackage{jheppub}	
\usepackage{graphicx}				
\usepackage{amsmath}
\usepackage{color}
\usepackage{mathtools}
\usepackage{amssymb}
\usepackage{booktabs}
\usepackage{comment}
\usepackage{mathrsfs}
\usepackage{enumerate}

\usepackage{hyperref}

\hypersetup{
    bookmarks=true,
    colorlinks,
    citecolor=blue,
    filecolor=black,
    linkcolor=blue,
    urlcolor=blue
}

\usepackage{overpic}
\usepackage[caption=false]{subfig}
\usepackage{array}
\usepackage{rotating}

\title{Action Complexity for Semi-Classical Black Holes}
                                           \author{Lukas Schneiderbauer,}
                                           \author{Watse Sybesma}
                                           \author{and L\'{a}rus Thorlacius}
                                           \affiliation{Science Institute,\\
                                           University of Iceland, \\Dunhaga 3, 107 Reykjav\'{i}k, Iceland.}
                                           \emailAdd{lukas.schneiderbauer@gmail.com}
                                           \emailAdd{watse@hi.is}
                                           \emailAdd{lth@hi.is}
\abstract{
We adapt the complexity as action prescription (CA) to a semi-classical model of 
two-dimensional dilaton gravity and determine the rate of increase of holographic complexity 
for an evaporating black hole. The results are consistent with our previous numerical results 
for semi-classical black hole complexity using a volume prescription (CV) in the same model, 
but the CA calculation is fully analytic and provides a non-trivial positive test for the holographic 
representation of the black hole interior. 
 }

\begin{document}
\maketitle

\section{Introduction}

The quantum complexity of a black hole is generated by the scrambling dynamics of quantum 
mechanical degrees of freedom that are enumerated by the black hole entropy.
These degrees of freedom can be usefully modelled in terms of a quantum circuit 
with k-local gates acting on a finite number of qubits.\footnote{For a recent review 
of complexity and  black holes, see \cite{Susskind:2018pmk}.}
In line with black hole complementarity \cite{Susskind:1993if}, the qubits can be taken to be 
located at (or near) a stretched horizon just outside the event horizon and the black hole 
interior is then viewed as an emergent spacetime region that provides a dual geometric 
description of the stretched horizon dynamics. 
In particular, the expanding spatial volume of the Einstein-Rosen bridge of a two-sided
eternal black hole is conjectured to reflect the growing complexity of the corresponding
quantum state \cite{Susskind:2014rva}. A refined version of the conjecture instead relates the 
complexity to the gravitational action evaluated on a specific bounded region of the 
black hole spacetime, referred to as the Wheeler-DeWitt (WdW) patch, that intersects 
the black hole interior \cite{Brown:2015bva,Brown:2015lvg}. 

The volume (CV) and action (CA) complexity conjectures have been explored for a 
variety of black hole geometries in Einstein gravity as well as extended to black hole solutions
in other gravitational theories. Much of the attention and effort has been focused on 
stationary black hole solutions while there have been fewer studies of dynamical
black holes (see {\it e.g.} \cite{Stanford:2014jda,Susskind:2014jwa,Chapman:2018dem,Chapman:2018lsv}). 
Black holes are stable in classical gravity and at late times, long after its formation by gravitational 
collapse, the geometry of a dynamical black hole will closely resemble that of a stationary one. 
Similarly, in the late time limit, the rate of growth of the holographic complexity (both CV and CA) 
of a classical dynamical black hole reduces to that of a stationary black hole of the same mass 
and other conserved charges (up to a factor of two accounting for the two-sided nature of the 
maximally extended stationary solution). This changes, however, when semi-classical effects 
are taken into account. In asymptotically flat spacetime, black hole evaporation due to Hawking 
emission results in the steady reduction of the black hole area and ends in a final state where 
there is an outgoing train of Hawking radiation but no black hole. 
For a large initial black hole mass, the quantum complexity of this final state 
will be very large, but finite, and presumably no longer growing as the Hawking radiation free
streams outwards \cite{Susskind:2018pmk}. 
 
In \cite{Schneiderbauer:2019anh} we initiated the study of holographic complexity in semi-classical 
gravity with the aim of testing the geometric representation of quantum complexity
in the context of black hole evaporation. In order to have analytical control, we considered semi-classical
toy models, the CGHS and RST models of two-dimensional dilaton gravity \cite{Callan:1992rs,Russo:1992ax}, 
that arise from the near-horizon limit of a near-extremal charged black hole in higher dimensions. 
We found non-trivial agreement between the volume 
of certain extremal surfaces and the expected behaviour of holographic complexity of classical
CGHS black holes and evaporating RST black holes, respectively. 
In \cite{Schneiderbauer:2019anh} we restricted our attention to volume complexity (CV), using a 
suitably defined volume functional that corresponds to spatial volume in the higher dimensional parent 
theory rather than geodesic length in the two-dimensional theory. For classical CGHS black holes, the 
volume complexity grows at a constant rate which is proportional to the product of 
Bekenstein-Hawking entropy and the Hawking temperature, 
\begin{equation}
\label{eq:firsteq}
\frac{d\mathcal{C}_V}{dt}\propto S\, T \,,
\end{equation}
as is expected on general grounds \cite{Susskind:2018pmk}. 
Here $t$ is the proper time of a distant fiducial observer and \eqref{eq:firsteq} holds 
both for two-sided eternal black holes and at late times for dynamical black holes formed by gravitational 
collapse. For semi-classical RST black holes, on the other hand, complexity growth slows down as
as the black hole evaporates and the rate of growth approaches zero at the endpoint of evaporation.
While the extremal volume could be obtained analytically for classical CGHS holes, we had to rely on 
numerical evaluation for semi-classical RST black holes in \cite{Schneiderbauer:2019anh}. Our 
numerical results confirmed that at leading order in a semi-classical expansion the rate of growth 
of the complexity, when expressed as a function of the proper time of a distant fiducial observer, 
is proportional to the product $S \,T$ for most of the black hole lifetime. 

In the present paper we extend this work by evaluating the holographic complexity
of semi-classical black holes in terms of an action on a Wheeler-DeWitt patch.
While this is more technically involved than the volume computations in \cite{Schneiderbauer:2019anh},
it has the distinct advantage that the entire semi-classical calculation can be carried out analytically
and yields explicit results for the rate of complexity growth of an evaporating RST black hole throughout its evolution. 
The relation \eqref{eq:firsteq} carries over to the semi-classical theory with the entropy at time $t$ given to leading 
order by the time-dependent area of the black hole. The next-to-leading order (logarithmic) term in 
the Bekenstein-Hawking entropy can also be read off from our analytic expression for the complexity growth and the
result agrees with previous semi-classical entropy calculations in the RST model \cite{Solodukhin:1995te,Myers1994}.
The numerical results for volume complexity obtained in our previous work \cite{Schneiderbauer:2019anh} are 
consistent with the new analytic results for action complexity. Together they provide a non-trivial positive test for 
black hole complementarity and the holographic duality between the stretched horizon and the black hole interior.

A Wheeler-DeWitt patch is bounded by co-dimension one and co-dimension two surfaces in spacetime. 
The associated gravitational action must include boundary terms in order to make the variational problem well-posed. 
For time-like and space-like co-dimension one boundaries the appropriate boundary terms in the two-dimensional theory 
are obtained from the standard Gibbons-Hawking-York term in the higher-dimensional parent theory, while contributions from 
null boundaries and co-dimension two boundaries require a more careful treatment \cite{Lehner:2016vdi}.
By working in so-called Kruskal gauge and arranging the boundary terms in the action to respect the same symmetry 
that simplifies the bulk field equations of the RST model \cite{Russo:1992ax}, we are able to eliminate a certain ambiguity
in the holographic action complexity and obtain a remarkably simple end result.

The structure of the paper is as follows. In Section~\ref{section:rst}, we carefully develop the boundary terms needed 
to have a well-posed variational problem for the two-dimensional dilaton gravity models. 
We then present the holographic action complexity (CA) for a classical CGHS black hole formed in gravitational 
collapse in Section~\ref{section:blackholes}, followed by the corresponding semi-classical calculation for an evaporating 
RST black hole in Section~\ref{section:blackholes_evap}. We end with a brief discussion and outlook for future work.

\section{RST model}\label{section:rst}

The action of the semi-classical RST model consists of three terms, 
\begin{equation}\label{eq:rst_model}
	S_\text{bulk}  = S_{0} +  S_{\text{q}} + S_{\text{ct}}
	\,,
\end{equation}
where
\begin{equation}\label{eq:CGHS}
	S_{0}
	=
	\int_{\mathcal{M}} d^{2}x\sqrt{-g}
	\left[
		e^{-2\phi}
		\left(
			R
			+
			4(\nabla\phi)^{2}
			+
			4\lambda^{2}
		\right)
		-
		\frac{1}{2}
		\sum^{N}_{i=i}(\nabla f_{i})^{2}
	\right]
\end{equation}
is the classical CGHS action involving a two-dimensional metric $g_{\mu \nu}$ along with a dilaton field $\phi$ 
and $N$ scalar matter fields $f_i$. The length scale $\lambda^{-1}$ is set by the magnetic charge of the  
higher-dimensional near extremal black hole and from now on we work in units where $\lambda=1$.
The second term,
\begin{equation}
	S_{\text{q}} =
	-\frac{\kappa}{4}\int d^{2}y\sqrt{-g}\left(R\frac{1}{\nabla^{2}}R\right)
	,
\end{equation}
with $\kappa=N/12$, was introduced by Callan {\it et al.}\ in \cite{Callan:1992rs} and captures the one-loop 
correction to the quantum effective action due to the conformal anomaly of the matter fields. For large $N$ this
term dominates over one-loop effects coming from the dilaton gravity sector that can therefore be ignored. 
The third term in the semi-classical action, 
\begin{equation}\label{eq:rst_ct}
	S_{\text{ct}} = -\frac{\kappa}{2}\int d^{2}y\sqrt{-g}\phi R
	\,,
\end{equation}
was introduced by Russo {\it et al.}\ in \cite{Russo:1992ax}. This term is allowed by general covariance and
does not disrupt the classical physics obtained in the limit $e^{-2\phi}\gg\kappa$. It enters at the same order as
$S_{\text{q}}$ and serves to preserve the classical symmetry of $S_{0}$ generated by the current 
$\partial^{\mu}(\rho-\phi)$, where $e^{2 \rho}$ is the conformal factor of the metric $g_{\mu\nu}$ with respect 
to a flat reference metric.
We set $\hbar=1$ throughout but note that when $\hbar$ is retained in the action it accompanies the
prefactor $\kappa$ and thus any expression involving $\kappa$ will be directly related to quantum corrections
in the semi-classical theory.

\subsection{A well-posed variational principle}

The CA proposal instructs us to evaluate the on-shell action of the model in question 
on a so-called WdW patch \cite{Brown:2015bva,Brown:2015lvg}.
However, it is well known that the action associated to a given set of equations of motion is not unique.
For instance, adding boundary terms does not change the equations of motion but will in general change the value of the action itself.
To restrict the set of possible actions, the CA proposal comes with the further requirement that the 
variational principle on the WdW patch should be well-posed. The equations of motion should 
follow from a stationary action principle assuming appropriate boundary conditions on the boundary of the WdW patch.
A solution to this problem was presented for Einstein-gravity in~\cite{Lehner:2016vdi}, 
where a particular set of co-dimension one boundary and co-dimension two joint terms were proposed.
In general, these terms are still not unique, but the requirements imposed in~\cite{Lehner:2016vdi} were enough 
to ensure a unique answer for the late-time complexity growth rate in well-known classical black hole geometries.
This is not the case, however, for the semi-classical model we consider below. Indeed, when we calculate the 
complexity growth for dynamical solutions that describe evaporating black holes, we find that certain boundary terms 
can be added that change the value of the action on the WdW patch while leaving the variational principle well-posed.
One therefore has to introduce further criteria, beyond those considered in~\cite{Lehner:2016vdi}, in order 
to have a definite prescription for the holographic action complexity. 
As laid out in the following, a sufficient criterion is to impose on the boundary terms the same symmetry 
that led to the simplification of the semi-classical field equations of the RST model itself in \cite{Russo:1992ax}.

To obtain a well-posed variational problem, we adapt the procedure proposed by~\cite{Lehner:2016vdi} to the situation at hand, 
a two-dimensional dilaton-gravity theory. However, a direct application is obstructed by the non-local term $S_{\text{q}}$. 
One way to remedy this problem is to introduce an auxiliary scalar field $Z$ and write the action in terms of integrals 
over local quantities only~\cite{Hayward:1994dw},
\begin{equation}\label{eq:rst_bulk}
	S_{\text{bulk}}
		=
		\int_{\mathcal{M}}d^{2}y\sqrt{-g}\left[R\tilde{\chi}+4\left(\left(\nabla\phi\right)^{2}+1\right)e^{-2\phi}-\frac{\kappa}{4}\left(\nabla Z\right)^{2}-\frac{1}{2}\sum_{i=1}^N\left(\nabla f_{i}\right)^{2}\right]
\end{equation}
with
\begin{equation}
	\tilde{\chi} := e^{-2\phi}-\frac{\kappa}{2}(\phi-Z)
	\, .
\end{equation}
As one can easily check, integrating out the auxiliary field $Z$ will return the original non-local action, up to boundary terms.

Considering a region with piecewise smooth space-like, time-like or null boundaries, the prescription of \cite{Lehner:2016vdi}
gives the following boundary terms involving the combination of fields $\tilde{\chi}$ that multiplies the Ricci scalar $R$ in the bulk
action,
\begin{equation}\label{eq:LMPS}
	\begin{aligned}
	S_\text{boundary}
	& =  2 \sum_{\mathcal{S}\in\mathscr{S}} \sigma_\mathcal{S} \int_\mathcal{S} \sqrt{|h|} \tilde{\chi} K d\Sigma \\
	&  + 2 \sum_{\mathcal{N}\in\mathscr{N}} \sigma_\mathcal{N} \int_\mathcal{N} d\lambda \tilde{\chi} \boldsymbol{\kappa}
	+ 2 \sum_{\mathcal{N}\in\mathscr{N}} \sigma_\mathcal{N} \int_{\mathcal{N}}d\lambda \, 
	\partial_{\lambda}\tilde{\chi}\log |\partial_{\lambda}f| \\ 
	&  + 2 \sum_{j\in\text{joints}} \sigma_j \tilde{\chi}\big|_j a_j
	\, .
	\end{aligned}
\end{equation}
The first term on the right hand side is the analogue of the familiar Gibbons-Hawking-York term, where $K$ is the extrinsic 
curvature of each time/space-like boundary component $\mathcal{S}\in\mathscr{S}$ and $h$ is the determinant of the induced 
metric on~$\mathcal{S}$.
The terms on the second line of \eqref{eq:LMPS} accompany null boundary components~$\mathcal{N}\in\mathscr{N}$. 
The integration variable $\lambda$ parametrizes the null line~$\mathcal{N}$. The failure of $\lambda$ to be an 
affine parameter is measured by~$\boldsymbol\kappa$, defined by the equation\footnote{Unfortunately, conventions dictate 
using the Greek letter kappa both in this context and as $\kappa=N/12$. We've opted for using boldface for one of them to reduce the
scope for confusion.}
\begin{equation}
	k^{\alpha} \nabla_\alpha k^{\beta} = {\boldsymbol \kappa} k^\beta
\end{equation}
with $k^\alpha := \frac{\partial{\gamma^\alpha}}{\partial{\lambda}}$ and $\gamma^\alpha(\lambda)$ being coordinates of 
the null curve $\mathcal{N}$ parametrized by $\lambda$.
The first term on the second line of~\eqref{eq:LMPS} is not invariant under reparametrizations $\lambda\mapsto\lambda^{\prime}$
by itself and the second term is added to offset this pathological feature.\footnote{See \cite{Lehner:2016vdi} for a detailed analysis.}
Here $f$ can a priori be an arbitrary function of any scalar field, provided $\partial_\lambda f$ does not vanish anywhere.

Finally, for each non-smooth joint $j$, we have to add a term $a_j$ which depends on the type and position of the joint. 
More explicitly, in the case of joints formed by two curves $\mathcal{S}_1$ and $\mathcal{S}_2$, that are separately either
spacelike or timelike, one finds
\begin{equation}
	a = \log \left| (n_1+p_1)_\mu n_2^\mu \right|
	,
\end{equation}
where $n_i$ are unit normal vectors to $\mathcal{S}_i$ and $p_1$ is a tangent vector of $\mathcal{S}_1$ that points 
outwards from the region of interest.

In case of a joint of two null-lines parametrized by $\lambda$ and $\bar{\lambda}$ respectively (and corresponding vectors 
$k^\alpha$ and $\bar{k}^\alpha$), $a$ reads
\begin{equation}
	a = \log \Big|  \frac{1}{2} k_\mu \bar{k}^\mu \Big|
	,
\end{equation}
while in case of a joint between a null and a space- or timelike boundary component, we have
\begin{equation}
	a = \log \left| k^\mu n_\mu \right|
\end{equation}
where $k^\mu$ corresponds to the null boundary in the way explained above and $n^\mu$ is the unit normal vector 
associated to the space- or timelike boundary.

The various terms in \eqref{eq:LMPS} are accompanied by signs $\sigma_\mathcal{S}$, $\sigma_\mathcal{N}$ and $\sigma_j$ 
that are sensitive to the conventions adapted in the procedure. A coherent set of rules is presented in~\cite{Lehner:2016vdi}.

Now, consider adding to the action a boundary term of the form
\begin{equation}\label{eq:example_term}
	\int_\mathcal{S} \sqrt{|h|} g(\phi,Z) d\Sigma
\end{equation}
involving an arbitrary function $g(\phi,Z)$. Adding a boundary term does not alter the equations of motion and 
a term of this particular form will not influence the variational principle if we impose Dirichlet boundary
conditions, {\it i.e.} take the variation of the induced metric $h$ and the variation of the scalar fields $\phi$ and $Z$ to 
vanish at the boundary. However, it is easy to see that such a term can drastically change the result of holographic complexity 
in our setup. Furthermore, considering regions with null-boundaries, \eqref{eq:LMPS} depends on an undetermined function $f$, 
which again influences the holographic complexity.
The above prescription thus needs to be supplemented by additional restrictions, as discussed below.

\subsection{RST symmetry}

In order to overcome the troublesome arbitrariness in the choice of boundary terms, we propose to restrict the allowed terms by an 
invariance requirement of the total action $S$ under the symmetry which guided the definition of the RST model in the first place.

In the following, we work in conformal gauge, where the line element takes the form
\begin{equation}\label{eq:conformal_gauge}
	ds^2 = - e^{2\rho} dy^{+} dy^{-}
	\, .
\end{equation}
Recall that the term $S_{\text{ct}}$, given by~\eqref{eq:rst_ct}, was introduced to preserve the symmetry generated by the current $\partial^{\mu}(\rho-\phi)$. The corresponding infinitesimal transformation of the fields $\phi$ and $\rho$ are given by~\cite{Cruz:1996zy,Russo:1992ax}
\begin{equation}
	\delta_{\scriptscriptstyle\text{RST}\,} \phi
	=
	\delta_{\scriptscriptstyle\text{RST}\,} \rho
	=
	\frac12 \delta_{\scriptscriptstyle\text{RST}\,} Z
	=
	\epsilon\, \frac{e^{2\phi}}{1-\frac{\kappa}{4}e^{2\phi}}
	\, ,
\end{equation}
while the matter fields do not transform.

We now impose the additional requirement that the total action $S$, \emph{including} boundary terms, remains 
invariant under the RST transformation, 
\begin{equation}
	\delta_{\scriptscriptstyle\text{RST}\,} S = 0
	\, .
\end{equation}
The bulk action \eqref{eq:rst_bulk} is invariant under $\delta_{\scriptscriptstyle\text{RST}}$ up to a boundary term 
that will have to be cancelled. 
Going back to the example~\eqref{eq:example_term}, it is evident that, generically, the RST variation of 
such a term will not vanish. We can use this to our advantage and choose the additional boundary term so that its 
variation cancels against the variation of the bulk action.

In order to work out the RST variation of the action $S_\text{bulk}+S_\text{boundary}$, it is convenient to define 
the fields\footnote{Note that this definition differs slightly from the one employed in \cite{Schneiderbauer:2019anh}.}
\begin{equation}\label{eq:def_omega_chi}
\begin{aligned}
	\Omega & :=e^{-2\phi}+\frac{\kappa}{2}\phi
	\, ,
	\\
	\chi & :=e^{-2\phi}+\kappa\rho-\frac{\kappa}{2}\phi
	\, ,
\end{aligned}
\end{equation}
for which the bulk action~\eqref{eq:rst_bulk} can be written as
\begin{equation}\label{eq:rst_bulk_rewrite}
	\begin{aligned}
	S_{\text{bulk}} = & \int d^2 y \sqrt{-g}\left[\frac{1}{\kappa} \left(\nabla\chi\right)^{2}- \frac{1}{\kappa}\left(\nabla\Omega\right)^{2}+\frac{2}{\sqrt{-g}}e^{\frac{2}{\kappa}(\chi-\Omega)}
	-\frac{1}{2}\sum_{i=1}^N\left(\nabla f_{i}\right)^{2}
	\right]
	\\
	& -2 \int d^2 y \sqrt{-g}\left[\nabla\left(\tilde{\chi}\nabla\rho\right)+\frac{\kappa}{8}\nabla\left(\eta\nabla\eta\right)\right] \,,
	\end{aligned}
\end{equation}
where $\eta$ is a harmonic field, $\nabla^2 \eta = 0$, obtained from the auxiliary $Z$ field via $Z=2\rho + \eta$.

The variations of the new fields $\Omega$ and $\chi$ can easily be evaluated and yield
\begin{equation}\label{eq:rst_var}
	\delta_{\scriptscriptstyle\text{RST}\,} \Omega = 
	\delta_{\scriptscriptstyle\text{RST}\,} \chi =
	-2 \epsilon 
	\, ,
\end{equation}
while $\delta_{\scriptscriptstyle\text{RST}\,}\eta=0$.
It is now evident that the RST variation of the first line of~\eqref{eq:rst_bulk_rewrite} vanishes and also
the variation of the last term on the second line. The remaining non-vanishing RST variation of the first 
total derivative term cancels against a contribution coming from the Gibbons-Hawking-York (GHY) term
that we consider next.

\subsubsection{Gibbons-Hawking-York boundary terms}
Let us now consider the GHY term in~\eqref{eq:LMPS} of the form
\begin{equation}
	2 \sigma_\mathcal{S} \int_\mathcal{S} \sqrt{|h|} \tilde{\chi} K d\Sigma
	\, ,
\end{equation}
with $K=\nabla_\mu n^\mu$ where $n^\mu$ is a unit normal vector to the surface $\mathcal{S}$.

In conformal gauge~\eqref{eq:conformal_gauge}, the following identity
\begin{equation}
	\sqrt{|h|}\nabla_{\mu}n^{\mu}
		=
		\sqrt{|h_{0}|}\partial_{\mu}n_{0}^{\mu}
		+\sqrt{|h|}n^{\mu}\partial_{\mu}\rho
\end{equation}
holds, where quantities with subscript $0$ are to be evaluated with respect to the flat reference metric 
$ds^2 = - dy^+ dy^-$. This implies that the GHY boundary term can be expressed as
\begin{equation}\label{eq:twoterms}
	2 \sigma_\mathcal{S} \int_\mathcal{S} \sqrt{|h|} \tilde{\chi} K \, d\Sigma
	=
	2 \sigma_\mathcal{S} \int_\mathcal{S} \sqrt{|h_0|} \tilde{\chi} K_0 \, d\Sigma
	+
	2 \sigma_\mathcal{S} \int_\mathcal{S} \sqrt{|h|} \tilde{\chi} n^{\mu} \partial_\mu \rho \, d\Sigma
	\, ,
\end{equation}
and furthermore, due to Stokes' theorem (assuming for the moment a region without null boundaries), 
the second term can be written as
\begin{equation}
	2 \sum_{\mathcal{S}\in{\mathscr{S}}} \sigma_\mathcal{S} \int_\mathcal{S} \sqrt{|h|} \tilde{\chi} n^{\mu} \partial_\mu \rho \, d\Sigma
	= 2 \int_\mathcal{M} d^2 y \sqrt{-g}\nabla(\tilde{\chi} \nabla \rho)
	\, ,
\end{equation}
which precisely cancels the first term in the second line of the bulk action~\eqref{eq:rst_bulk_rewrite}. 
This leaves us only with the term involving the flat reference metric in \eqref{eq:twoterms}. This term
does not contribute when we obtain field equations using field variations (under which the flat reference metric is fixed).
However, its RST variation does not vanish in general, 
\begin{equation}
	\delta_{\text{RST}} \left(
		2\sum_{\mathcal{S}} \sigma_\mathcal{S} \int_\mathcal{S} \sqrt{|h_0|} \tilde{\chi} K_0 \, d\Sigma
	\right)
	=
	- 4 \epsilon \sum_{\mathcal{S}} \sigma_\mathcal{S} \int_\mathcal{S} \sqrt{|h_0|} K_0 \, d\Sigma
	\neq 0 \,.
\end{equation}
This variation can be cancelled by introducing a suitably chosen additional boundary term.

\subsubsection{Time-/spacelike joint contributions}
Still assuming no null boundaries, this leaves us with the analysis of the joint contributions to the 
action~$S_\text{boundary}$ in~\eqref{eq:LMPS}, i.e.
\begin{equation}
	2 \sum_{j\in\text{joints}} \sigma_j \tilde{\chi}\big|_j
	\log|(n^j_1+p^j_1)_\mu (n^j_2)^\mu|
	\, .
\end{equation}
It turns out that in two dimensions, these terms are actually not necessary to obtain a well-posed variational principle. 
The reason is that the argument of the logarithm does not depend on the conformal factor at all. 
This is easily seen by writing the unit normal vector as
\begin{equation}\label{eq:normal_v}
	n_\mu
	=
	\frac{\partial_\mu \Phi}{ \sqrt{ g^{\sigma \rho} \partial_\sigma \Phi \partial_\rho \Phi} }
	=
	e^\rho \frac{\partial_\mu \Phi}{ \sqrt{ g_0^{\sigma \rho} \partial_\sigma \Phi \partial_\rho \Phi} }
	=
	e^\rho (n_0)_\mu
\end{equation}
where $\Phi$ is temporarily introduced as a scalar field whose contour lines describe the surface $\mathcal{S}$ locally. 
Similarly,
\begin{equation}
	n^{\mu} = e^{-\rho} n_0^{\mu}
\end{equation}
and the same is true for the tangent vector $p_1$. It follows that the inner product is independent of the conformal factor~$\rho$.

Since the term is proportional to $\tilde{\chi}$, its RST variation is easily evaluated,
\begin{equation}
	\delta_{\text{RST}} \left( 2 \sigma_j \tilde{\chi}\big|_j a_j \right)
	=
	- 4\epsilon \, \sigma_j a_j,
\end{equation}
which does not vanish in general. The RST symmetry is easily enforced by simply leaving out joint terms of this form. 
This is possible, because, as we have just seen, such terms do not influence the variational principle in two dimensions.

\subsubsection{Null boundary contributions}
Let us now include null boundaries in our discussion. It will be beneficial to rewrite the terms in~\eqref{eq:LMPS} corresponding to null boundaries,
\begin{equation}\label{eq:null_contr}
	2 \sigma_\mathcal{N} \int_\mathcal{N} d\lambda \tilde{\chi} \boldsymbol{\kappa}
	+ 2 \sigma_\mathcal{N} \int_{\mathcal{N}}d\lambda \, \partial_{\lambda}\tilde{\chi}\log |\partial_{\lambda}f|
	,
\end{equation}
in a way that is manifestly reparametrization invariant.
This is achieved by integrating the second term by parts which gives
\begin{equation}
	2 \sigma_\mathcal{N} \int_{\mathcal{N}}d \lambda\partial_{\lambda}\tilde{\chi}\log |\partial_{\lambda}f|
	=
	2 \sigma_\mathcal{N} \tilde{\chi} \log |\partial_{\lambda}f|\bigg|_{1}^{2}
	-2 \sigma_\mathcal{N} \int_{\mathcal{N}}d\lambda\tilde{\chi}\partial_{\lambda}\log|\partial_{\lambda}f|
\end{equation}
and using that
\begin{equation}
	\partial_{\lambda}\log|\partial_{\lambda}f|
	= \boldsymbol{\kappa} +
	\frac{k_{-}^{\mu}k_{-}^{\nu}\nabla_{\mu}\partial_{\nu}f}{k_{-}^{\sigma}\partial_{\sigma}f}
	\, ,
\end{equation}
we note, that the term involving $\boldsymbol{\kappa}$ cancels the original $\boldsymbol{\kappa}$-dependent 
term in~\eqref{eq:null_contr}, so that in total we have
\begin{equation}\label{eq:null_contr_rewr}
	\text{\eqref{eq:null_contr}} =
	2 \sigma_\mathcal{N} \tilde{\chi} \log |\partial_{\lambda}f|\bigg|_{1}^{2}
	-2 \sigma_\mathcal{N} \int_{\mathcal{N}}d\lambda \tilde{\chi} 
	\frac{k_{-}^{\mu}k_{-}^{\nu}\nabla_{\mu}\partial_{\nu}f}{k_{-}^{\sigma}\partial_{\sigma}f}
	\, .
\end{equation}
The second term is now manifestly invariant under a change of parametrization $\lambda\mapsto\lambda^\prime:=e^\beta \lambda$, 
since $k^\alpha\mapsto e^{-\beta} k^\alpha$. When the original joint terms are combined with the new terms obtained from
integration by parts in~\eqref{eq:null_contr_rewr} the full expression is also reparametrization invariant. Independent of the precise 
nature of the joints, the original contribution will be of the form
\begin{equation}\label{eq:joint_general}
	2 \sigma \tilde{\chi} \log | A \, m_\mu k^\mu |
	,
\end{equation}
where $A$ is a constant, $k^\alpha$ is the null vector associated with the null boundary in question, and $m^\mu$ is a vector 
that depends on nature of the joint, which we will be agnostic about for this argument. 
The sign~$\sigma$ depends on conventions, but the relative sign to~$\sigma_\mathcal{N}$ is fixed by 
\begin{equation}
	\sigma=\begin{cases}
		-\sigma_{\mathcal{N}} & \text{if joint lies in the future of \ensuremath{\mathcal{N}}}\\
		+\sigma_{\mathcal{N}} & \text{if joint lies in the past of \ensuremath{\mathcal{N}}}
	\end{cases}
\end{equation}
assuming $k^\mu$ is future directed. This implies, that for the total joint contributions on either side of the null boundary, we obtain
\begin{equation}\label{eq:joint_repr_inv}
	\text{\eqref{eq:joint_general}} - 2 \sigma \tilde{\chi} \log |\partial_{\lambda}f|
	= 2 \sigma \tilde{\chi} \log \Big| A \, \frac{m_\mu k^\mu}{\partial_\lambda f} \Big|
	,
\end{equation}
which is also manifestly invariant under reparametrization $\lambda\mapsto\lambda^\prime$.

We have now successfully rewritten the terms corresponding to a null boundary and its attached joints in a 
manifestly reparametrization invariant form.
Next, in order to obtain the RST variation, it will again be convenient to split those terms into parts which depend on the 
conformal factor and parts which do not. We have
\begin{equation}
	-2 \sigma_\mathcal{N} \int_{\mathcal{N}}d\lambda \tilde{\chi} 
	\frac{k_{-}^{\mu}k_{-}^{\nu}\nabla_{\mu}\partial_{\nu}f}{k_{-}^{\sigma}\partial_{\sigma}f}
	=
	-2 \sigma_\mathcal{N} \int_{\mathcal{N}}d\lambda \tilde{\chi} \frac{k_{-}^{\mu}k_{-}^{\nu}\partial_{\mu}\partial_{\nu}f}{k_{-}^{\sigma}\partial_{\sigma}f}
	+4 \sigma_\mathcal{N} \int_{\mathcal{N}}d\lambda \tilde{\chi} \partial_\lambda \rho
	\, ,
\end{equation}
where the first term is reparametrization invariant and does not depend on the conformal factor $\rho$.

Considering the joint terms~\eqref{eq:joint_repr_inv}, we can write them as 
\begin{equation}
	\begin{aligned}
	& - 2 \sigma_\mathcal{N} \tilde{\chi}_2 \log \Big| A \, \frac{m_\mu k^\mu}{\partial_\lambda f} \Big|_2
	+
	2 \sigma_\mathcal{N} \tilde{\chi}_1 \log \Big| \bar{A} \, \frac{\bar{m}_\mu k^\mu}{\partial_\lambda f} \Big|_1 \\
	= &
	- 2 \sigma_\mathcal{N} \int_\mathcal{N} d\lambda \partial_\lambda \left( \tilde{\chi} \log(e^{\rho}) \right)
	- 2 \sigma_\mathcal{N} \tilde{\chi}_2 \log \Big| A \, \frac{e^{-\rho} m_\mu k^\mu}{\partial_\lambda f} \Big|_2
	+
	2 \sigma_\mathcal{N} \tilde{\chi}_1 \log \Big| \bar{A} \, \frac{e^{-\rho} \bar{m}_\mu k^\mu}{\partial_\lambda f} \Big|_1
	\, .
	\end{aligned}
\end{equation}
One can easily check that $e^{-\rho} m_{\mu}$ does not depend on $\rho$ in case of a joint with a space- or timelike curve 
(since then $m^\mu$ is given by a normal vector, see~\eqref{eq:normal_v}). In case of a joint between two null curves, 
the argument of the logarithm is given by $g_{\mu\nu}\bar{k}^\mu k^\nu=e^{2\rho} \eta_{\mu\nu}\bar{k}^\mu k^\nu$, 
but now the above procedure is performed twice (once for each null surface), so that the resulting argument is 
$\eta_{\mu\nu}\bar{k}^\mu k^\nu$. Hence, in all cases, the resulting joint terms will be independent of the conformal factor~$\rho$.

Combining the resulting terms, and using the product rule, we obtain
\begin{equation}\label{eq:null_remaining_terms}
	\begin{aligned}
	-2 \sigma_\mathcal{N} \int_{\mathcal{N}}d\lambda \tilde{\chi} \frac{k_{-}^{\mu}k_{-}^{\nu}\partial_{\mu}\partial_{\nu}f}{k_{-}^{\sigma}\partial_{\sigma}f}
	- 2 \sigma_\mathcal{N} \tilde{\chi}_2 \log \Big| A \, \frac{e^{-\rho} m_\mu k^\mu}{\partial_\lambda f} \Big|_2
	+
	2 \sigma_\mathcal{N} \tilde{\chi}_1 \log \Big| \bar{A} \, \frac{e^{-\rho} \bar{m}_\mu k^\mu}{\partial_\lambda f} \Big|_1	 
	\\
	+2 \sigma_\mathcal{N} \int_{\mathcal{N}}d\lambda \tilde{\chi} \partial_\lambda \rho
	- 2 \sigma_\mathcal{N} \int_\mathcal{N} d\lambda \rho \partial_\lambda \tilde{\chi}
	\, ,
	\end{aligned}
\end{equation}
where each term is now manifestly invariant under reparametrization~$\lambda\mapsto\lambda^\prime$ and the first line 
is independent of the conformal factor~$\rho$.
Importantly, the first term in the second line combines with the corresponding terms arising from space- or timelike boundary 
components, in conjunction with Stokes' theorem,\footnote{See appendix~\ref{section:stoke_null} for a justification of the 
formula when including null boundaries.}
\begin{equation}
	2 \sum_{\mathcal{N}\in\mathscr{N}} \sigma_\mathcal{N} \int_{\mathcal{N}}d\lambda \tilde{\chi} \partial_\lambda \rho
	+ 2 \sum_{\mathcal{S}\in{\mathscr{S}}} \sigma_\mathcal{S} \int_\mathcal{S} \sqrt{|h|} \tilde{\chi} n^{\mu} \partial_\mu \rho \, d\Sigma
	= 2 \int_\mathcal{M} d^2 y \sqrt{-g}\nabla(\tilde{\chi} \nabla \rho)
	\, ,
\end{equation}
in order to cancel with the total derivative contribution coming from the bulk~\eqref{eq:rst_bulk_rewrite}. 

Since the RST variation of the remaining bulk contribution vanishes, the RST variation of the null contributions has to vanish as well. 
Because the first line of~\eqref{eq:null_remaining_terms} is independent of the conformal factor, these terms are not necessary 
in order to ensure a well-posed variational problem and we leave them out to implement the RST symmetry. 
As these terms are manifestly reparametrization invariant, removing them will not spoil overall reparametrization invariance.
The RST variation of the last term in the last line of~\eqref{eq:null_remaining_terms} does not vanish, but since it is 
reparametrization invariant and the variation of $\rho$ vanishes at the boundary (there are no derivatives involved),
it can be cancelled by adding a boundary term.

\subsubsection{Complete action}
Let us now collect the results of the above considerations. We obtain a simple expression for the total action~$S$, in conformal gauge,
\begin{equation}
	\begin{aligned}
		S & = \int_\mathcal{M} d^2 y \sqrt{-g}\left[\frac{1}{\kappa} \left(\nabla\chi\right)^{2}- \frac{1}{\kappa}\left(\nabla\Omega\right)^{2}+\frac{2}{\sqrt{-g}}e^{\frac{2}{\kappa}(\chi-\Omega)}
		-\frac{1}{2}\sum_{i=1}^N\left(\nabla f_{i}\right)^{2}
		\right]
		\\
		& - \frac{\kappa}{4} \int d^2 y \sqrt{-g}\left[\nabla\left(\eta\nabla\eta\right)\right]
		\, .
	\end{aligned}
\end{equation}
This action has the properties that the variational principle is well-posed on any spacetime region bounded by spacelike, timelike, 
or null boundaries. Additionally, it is invariant under RST transformations in the sense that $\delta_{\scriptscriptstyle\text{RST}}S = 0$.
Note that the action does not involve any boundary or joint terms anymore, since they were either consistently removed, 
or canceled against total derivative contributions from the original bulk action.

As a result, the expression for the holographic complexity does not involve an arbitrary function $f$ anymore. 
Further, as a side note, the value of holographic complexity on the WdW patch can also be obtained by a limiting procedure, 
regulating the WdW patch with space- or timelike surfaces only. The resulting limit is finite, and it agrees with the result 
obtained by the above RST symmetric prescription for null boundaries.

\paragraph{The on-shell action}
The equation of motion
\begin{equation}
	\nabla^2 \chi = \nabla^2\Omega
\end{equation}
allows us to choose Kruskal coordinates $(x^{+},x^{-})$ where $\rho=\phi$, implying $\Omega=\chi$. 
In this coordinate system, the on-shell action~$S$ is subject to a remarkable simplification, 
\begin{equation}\label{eq:total_onshell_action}
	S = \int_\mathcal{M} dx^{+} dx^{-}
	\left[ 2 
		-\sum_{i=1}^N \partial_{+} f_{i} \partial_{-} f_{i}
		+ \frac{\kappa}{2} \partial_{+} \eta \partial_{-} \eta
	\right]
	\, .
\end{equation}
In particular, in this form the action has no explicit dependence on the the dilaton field~$\phi$. This is, of course, somewhat 
misleading, for the shape of the WdW patch in Kruskal coordinates will indeed depend on the spacetime metric and therefore 
the dilaton as well.

\section{Classical Black Hole Complexity}\label{section:blackholes}

\subsection{Gravitational Collapse}

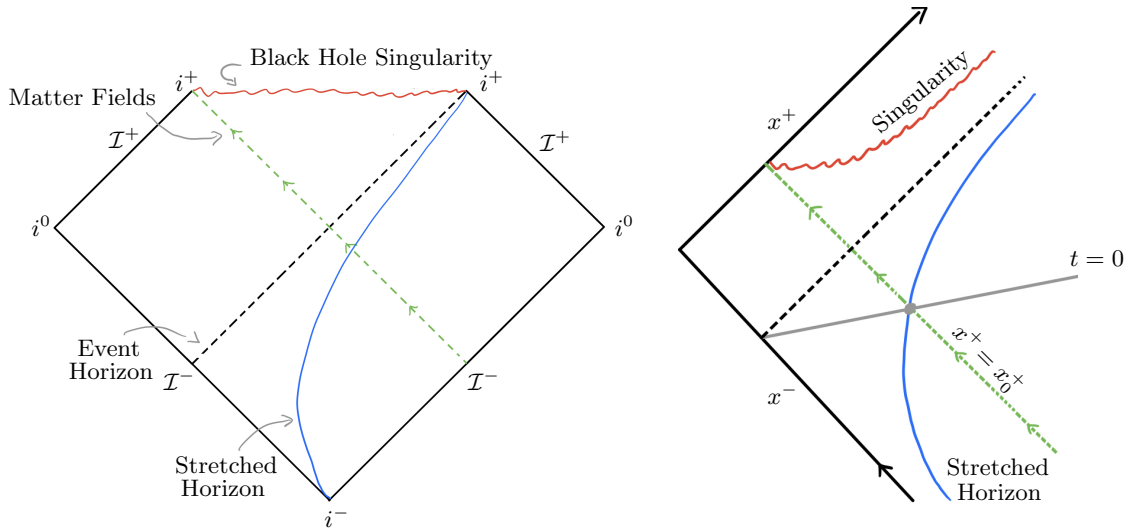
\begin{figure}[h]
	\begin{center}
		\begin{overpic}[width=.50\textwidth]{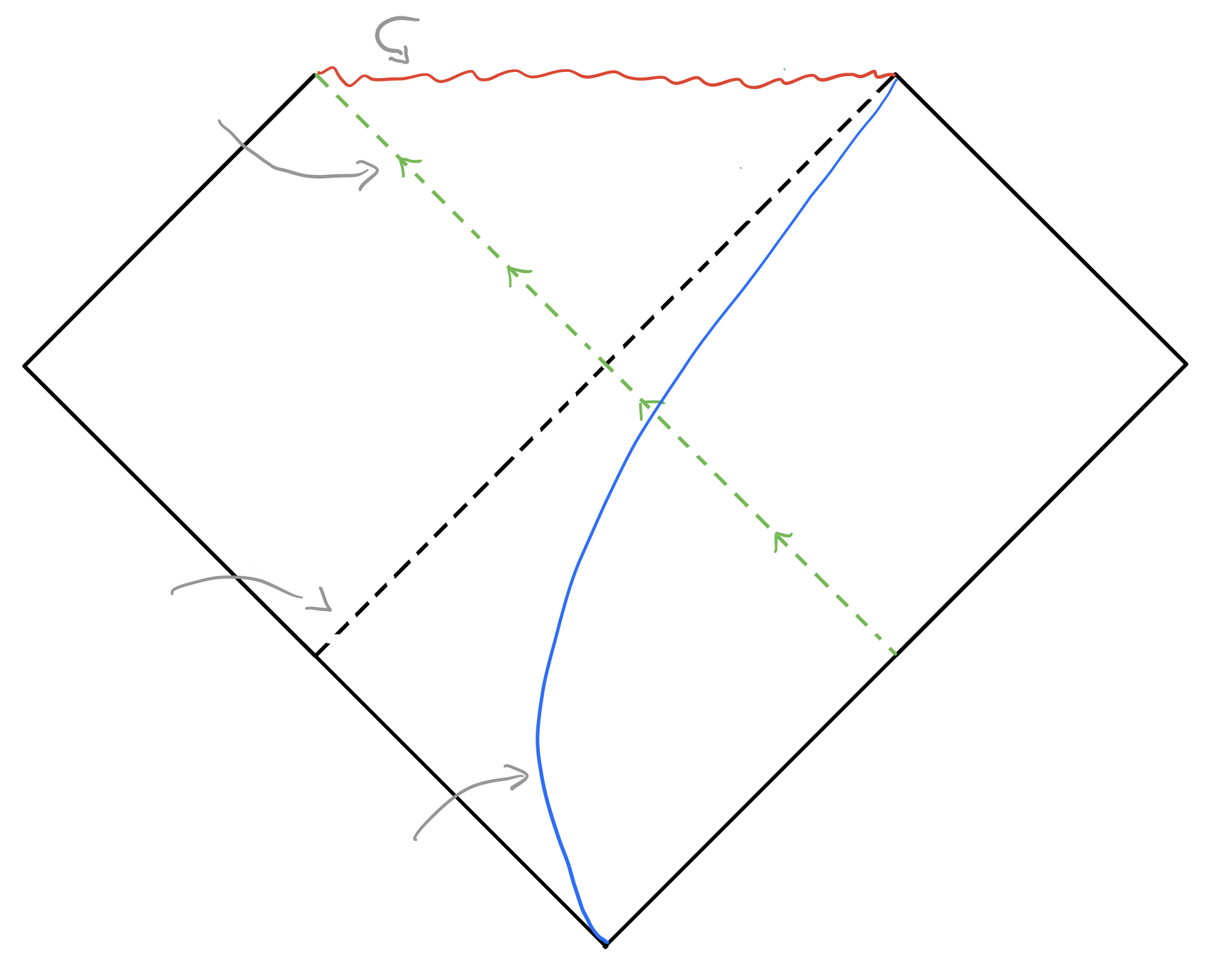}
			\put (-6,72) {\footnotesize{Matter Fields}}
			\put (36,79) {\footnotesize{Black Hole Singularity}}
			\put (23,7.5) {\footnotesize{Stretched}}
			\put (24,3.5) {\footnotesize{Horizon}}
			\put (6,28) {\footnotesize{Event}}
			\put (4.5,24) {\footnotesize{Horizon}}
			\put (21,22) {\footnotesize{$\mathcal{I}^{-}$}}
			\put (74,22) {\footnotesize{$\mathcal{I}^{-}$}}
			\put (86.5,64) {\footnotesize{$\mathcal{I}^{+}$}}
			\put (11.5,64.5) {\footnotesize{$\mathcal{I}^{+}$}}
			\put (-2,49) {\footnotesize{$i^{0}$}}
			\put (99.5,49) {\footnotesize{$i^{0}$}}
			\put (49,-2) {\footnotesize{$i^{-}$}}
			\put (23,74.5) {\footnotesize{$i^{+}$}}
			\put (76,74.5) {\footnotesize{$i^{+}$}}
		\end{overpic}
		\hspace{.5cm}
		\begin{overpic}[width=.35\textwidth]{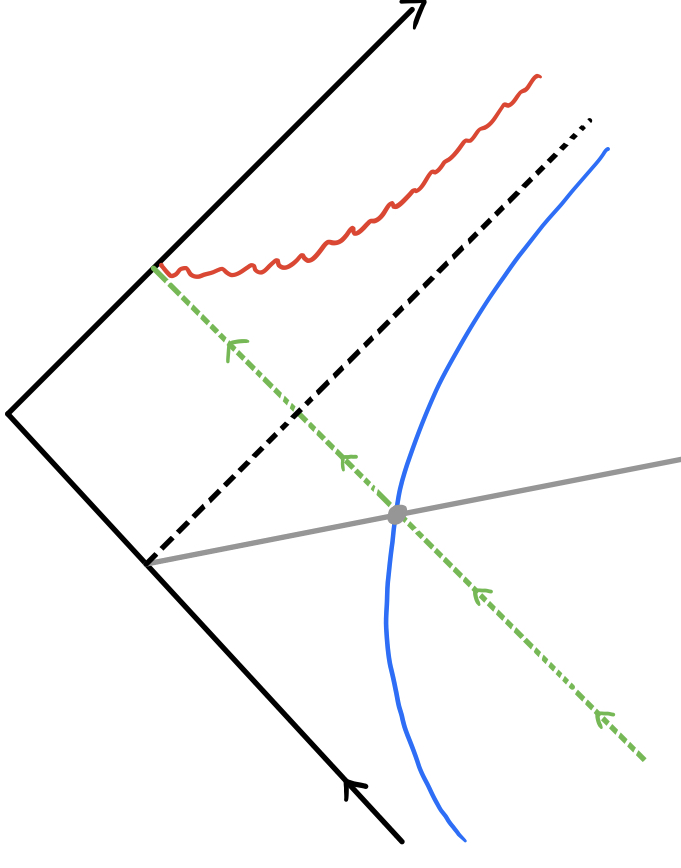}
			\put (18,22) {\footnotesize{$x^{-}$}}
			\put (18,76) {\footnotesize{$x^{+}$}}
			\put (53,36) {\rotatebox{-45}{\footnotesize{$x^{+}\!=x^{+}_{0}$}}}
			\put (38,75) {\rotatebox{40}{\footnotesize{Singularity}}}
			\put (53,8) {\footnotesize{Stretched}}
			\put (55,3) {\footnotesize{Horizon}}
			\put (77,49) {\footnotesize{$t=0$}}
		\end{overpic}
	\end{center}
\vspace{-0.3cm}
\caption{
Left panel: Penrose diagram of one-sided CGHS black hole formed by gravitational collapse. Right panel: The corresponding
Kruskal diagram with the same color coding. The gray line denotes a curve of equal tortoise time $t$.}
\label{fig:collapse_1}
\end{figure}

Before discussing the semi-classical case, let us analyse the classical gravitational collapse of an infinitely thin shell of 
incoming matter $f$ with mass $M$. The energy momentum tensor associated to the matter field $f$ is given by
\begin{equation}
	T^{f}_{++}
	=
	\frac{M}{x_{0}^{+}}\delta(x^{+}-x^{+}_{0})
	\,,
\end{equation}
and for the dilaton~$\phi$ and conformal factor $\rho$ this implies
\begin{equation}
	e^{-2\phi}
	=
	e^{-2\rho}
	=
	\begin{cases}
		-x^{+}x^{-}	& \text{if } x^{+}< x^{+}_{0}\\
		-x^{+}\left(x^{-}+\frac{M}{x_{0}^{+}}\right)+M
		&  \text{if } x^{+}\geq x^{+}_{0}
		\,,
	\end{cases}
\end{equation}
in Kruskal coordinates. The infalling shell creates a black hole singularity, as shown in Figure~\ref{fig:collapse_1}, 
which depicts a Penrose and Kruskal diagram on the left and right, respectively.

In line with our previous paper~\cite{Schneiderbauer:2019anh}, we take the WdW patch to be anchored at the stretched horizon, 
defined as a membrane outside the black hole, with an area that is one unit larger than the area of the event horizon,
\begin{equation}\label{eq:SHdef}
	e^{-2\phi_{\text{SH}}}=e^{-2\phi_{\text{EH}}}+1=M+1
	\,.
\end{equation}
In the classical collapse solution considered here, the stretched horizon is a curve of constant dilaton~$\phi$ outside the black hole.
If we instead anchor the WdW patch on a curve far outside the black hole, the main difference is to shift
the onset of complexity growth forward in time, to the time in tortoise coordinates when the infalling shock wave passes through 
the anchor curve on its way to forming the black hole. As was discussed in~\cite{Schneiderbauer:2019anh}, it seems more 
physical to place the anchor curve at the stretched horizon and have the onset of complexity growth coincide, at least 
approximately, with the time of black hole formation (here defined as the tortoise time at which the shock wave meets the 
stretched horizon). 

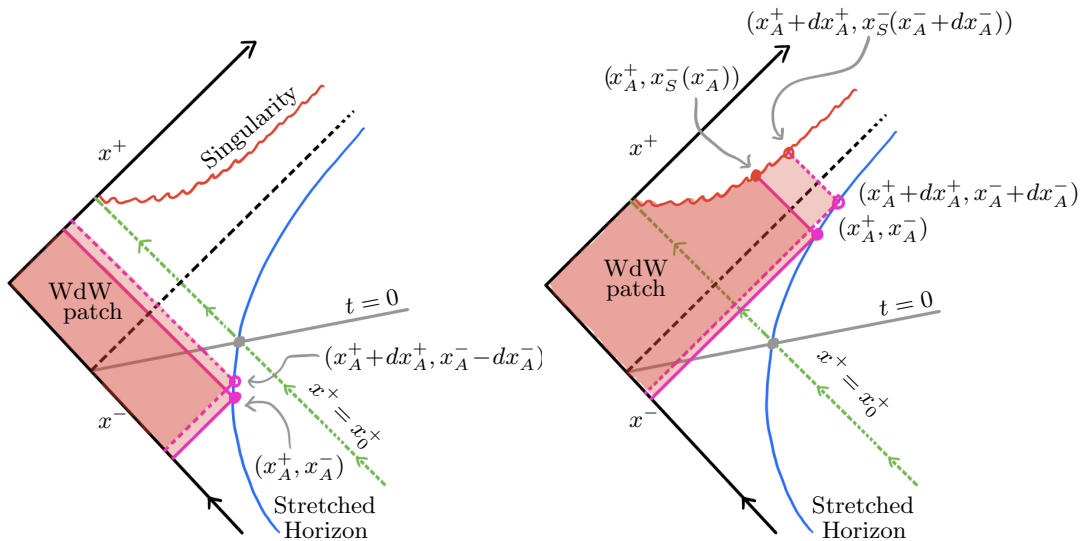
\begin{figure}[h]
	\vspace{0.5cm}
	\begin{center}
		\begin{overpic}[width=.35\textwidth]{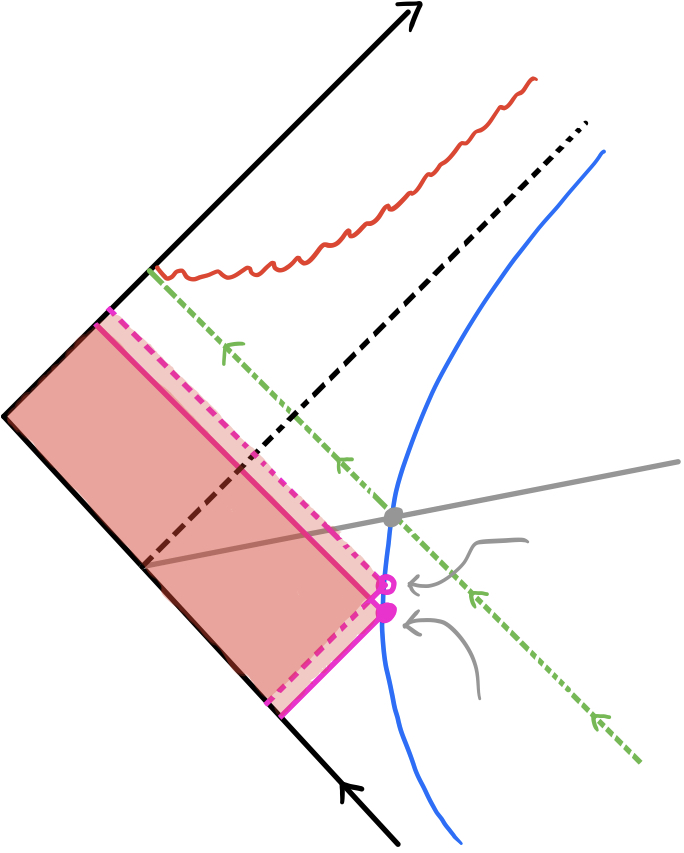}
			\put (8,48) {\footnotesize{WdW}}
			\put (11,44) {\footnotesize{patch}}
			\put (18,22) {\footnotesize{$x^{-}$}}
			\put (18,76) {\footnotesize{$x^{+}$}}
			\put (59,30) {\rotatebox{-45}{\footnotesize{$x^{+}\!=x^{+}_{0}$}}}
			\put (38,74) {\rotatebox{40}{\footnotesize{Singularity}}}
			\put (53,5) {\footnotesize{Stretched}}
			\put (55,0) {\footnotesize{Horizon}}
			\put (67,45) {\rotatebox{11}{\footnotesize{$t=0$}}}
			\put (49,13) {\footnotesize{$(x^{+}_{A},x^{-}_{A})$}}
			\put (63,35) {\footnotesize{$(x^{+}_{A}\!+\!dx^{+}_{A},x^{-}_{A}\!-\!dx^{-}_{A})$}}
		\end{overpic}
		\hspace{1.5cm}
		\begin{overpic}[width=.35\textwidth]{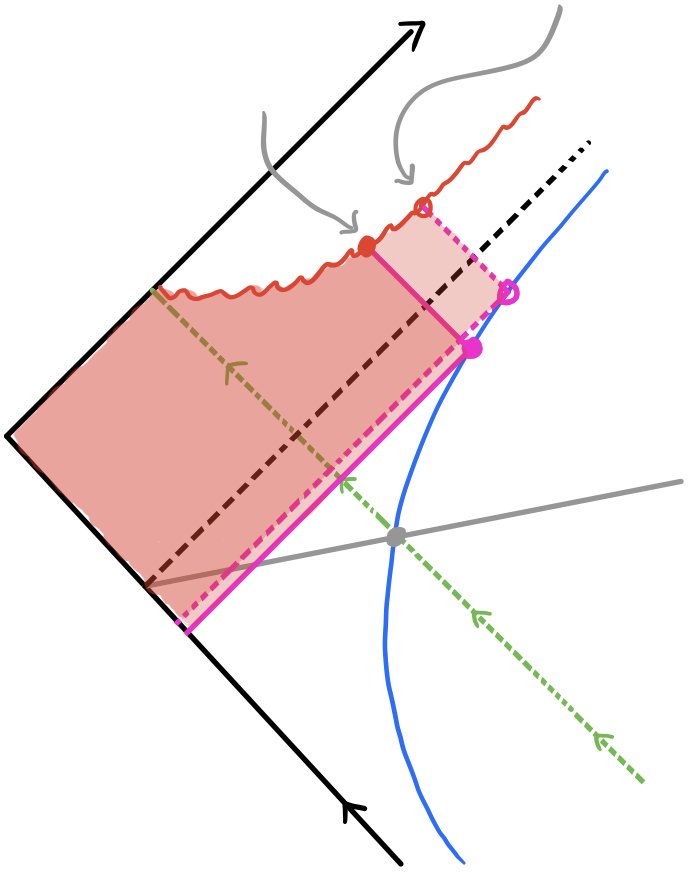}
			\put (12,52) {\footnotesize{WdW}}
			\put (12.5,48) {\footnotesize{patch}}
			\put (17,22) {\footnotesize{$x^{-}$}}
			\put (17,76) {\footnotesize{$x^{+}$}}
			\put (53,36) {\rotatebox{-45}{\footnotesize{$x^{+}\!=x^{+}_{0}$}}}
			\put (12,90) {\rotatebox{0}{\footnotesize{$(\!x^{+}_{A},x^{-}_{S}(x^{-}_{A}))$}}}
			\put (40,101) {\rotatebox{0}{\footnotesize{$(x^{+}_{A}\!+\!dx^{+}_{A},x^{-}_{S}\!(x^{-}_{A}\!+\!dx^{-}_{A}))$}}	}	
			\put (53,5) {\footnotesize{Stretched}}
			\put (55,0) {\footnotesize{Horizon}}
			\put (66,44) {\rotatebox{11}{\footnotesize{$t=0$}}}
			\put (58,60) {\footnotesize{$(x^{+}_{A},x^{-}_{A})$}}
			\put (62,67) {\footnotesize{$(x^{+}_{A}\!+\!dx^{+}_{A},x^{-}_{A}\!+\!dx^{-}_{A})$}}			
		\end{overpic}
	\end{center}
\caption{
Evolution of the WdW patch for classical gravitational collapse. Color coding coincides with Figure \ref{fig:collapse_1}.
}\label{fig:collapsewdw}
\end{figure}

The WdW patch at a given tortoise time $t$ is defined as the union of all spacelike surfaces originating from the point 
on the anchor curve that intersects the appropriate constant $t$ curve and extending towards the black hole, see 
Figure~\ref{fig:collapsewdw}. 
The holographic complexity $\mathcal{C}$ at time $t$ is then given by the action evaluated on the WdW patch.
The classical action can be obtained by formally setting $\kappa$ to zero in~\eqref{eq:total_onshell_action}.
Furthermore, the collapse solution only involves infalling matter, $\partial_{-} f=0$, and the on-shell 
action~\eqref{eq:total_onshell_action} reduces to
\begin{equation}\label{eq:total_onshell_action_cl}
	\mathcal{C} = 
	S_\text{cl} = 2 \int_\text{WdW} dx^{+} dx^{-} =: 2 \mathcal{A}
	\, ,
\end{equation}
where $\mathcal{A}$ can be interpreted as the reference metric ``area" of the WdW patch drawn in Kruskal coordinates.

An asymptotic observer would use the tortoise coordinates $(t,x)$, which are related to Kruskal coordinates $(x^{+},x^{-})$ 
by the equations
\begin{equation}\label{eq:shifted_tortoise}
	\begin{aligned}
	x^{+} & =e^{t+x}\,,\\
	x^{-}+\frac{M}{x^{+}_{0}} & =-e^{-t+x}.
	\end{aligned}
\end{equation}
We are interested in the growth rate of holographic complexity~$\frac{d\mathcal{C}}{dt}$, where $t$ is the tortoise time 
associated to the anchor point of the WdW patch, see Figure~\ref{fig:collapse_1}.
To this end, we denote the Kruskal coordinates describing the anchor point as $(x^{+}_A,x^{-}_A)$, while the singularity 
curve is described by $(x^{+}_S,x^{-}_S)$.
It is practical to consider two separate cases: the WdW patch anchored before the shock wave arrives, 
$x^{+} < x^{+}_0$, and after, $x^{+} > x^{+}_0$.

\paragraph{Before incoming shockwave}
It is apparent from Figure~\ref{fig:collapsewdw} that the change of the area in Kruskal coordinates, 
$\mathcal{A}$, before the shockwave arrives, is given by 
\begin{equation}
	d \mathcal{A} =
		-  x^{-}_A d x^{+}_A
		-  x^{+}_A d x^{-}_A
	\, .
\end{equation}
From the definition of the stretched horizon~\eqref{eq:SHdef}, which we identify with the anchor curve, 
we obtain the equivalent relation
\begin{equation}
	- x^{+}_A \left( x^{-}_A + \frac{M}{x^{+}_0} \right) = 1
	\, ,
\end{equation}
which immediately implies that
\begin{equation}\label{eq:anchor_curve_change}
	- x^{-}_A d x^{+}_A - x^{+}_A d x^{-}_A = \frac{M}{x^{+}_0} d x^{+}_A
	\, ,
\end{equation}
so that
\begin{equation}
	d \mathcal{A} = \frac{M}{x^{+}_0} d x^{+}_A
	\, .
\end{equation}
Furthermore, since $d x^{+}_A = x^{+}_A dt$, if we shift the time variable~$t$, so that $t=0$ corresponds to
where the shock wave meets the stretched horizon, then $d \mathcal{A} = M e^t dt$, or
\begin{equation}
	\dot{\mathcal{C}} = 2 M e^t	\quad \text{for } x^{+} < x^{+}_0
	\, .
\end{equation}
We observe an exponential onset towards $2M$ at $t=0$, the black hole creation time.

\paragraph{After incoming shockwave}
The analogous calculation can be done for times after the black hole creation, $t>0$. One easily finds
\begin{equation}
	d \mathcal{A} =
		\left( x^{-}_S(x^{+}_A) - x^{-}_A \right) d x^{+}_A
		-  x^{+}_A d x^{-}_A
	\, .
\end{equation}
Using~\eqref{eq:anchor_curve_change} in conjuction with the defining relation for the black hole singularity,
\begin{equation}
	M = x^{+}_S \left(x^{-}_S + \frac{M}{x^{+}_0}\right)
	\, ,
\end{equation}
one obtains
\begin{equation}
	d \mathcal{A}
	= \left( x^{-}_S(x^{+}_A) + \frac{M}{x^{+}_0} \right) d x^{+}_A
	=  \frac{M}{x^{+}_A} d x^{+}_A
	= M dt
	\, ,
\end{equation}
or
\begin{equation}
	\dot{\mathcal{C}} = 2 M \quad \text{for } x^{+} > x^{+}_0
	\, ,
\end{equation}
so that the holographic complexity growth~$\dot{\mathcal{C}}$ is continuous at~$x^{+}=x^{+}_0$ and remains
constant for $x^{+}>x^{+}_0$.

\begin{figure}
	\begin{center}
		\begin{overpic}[width=.6\textwidth]{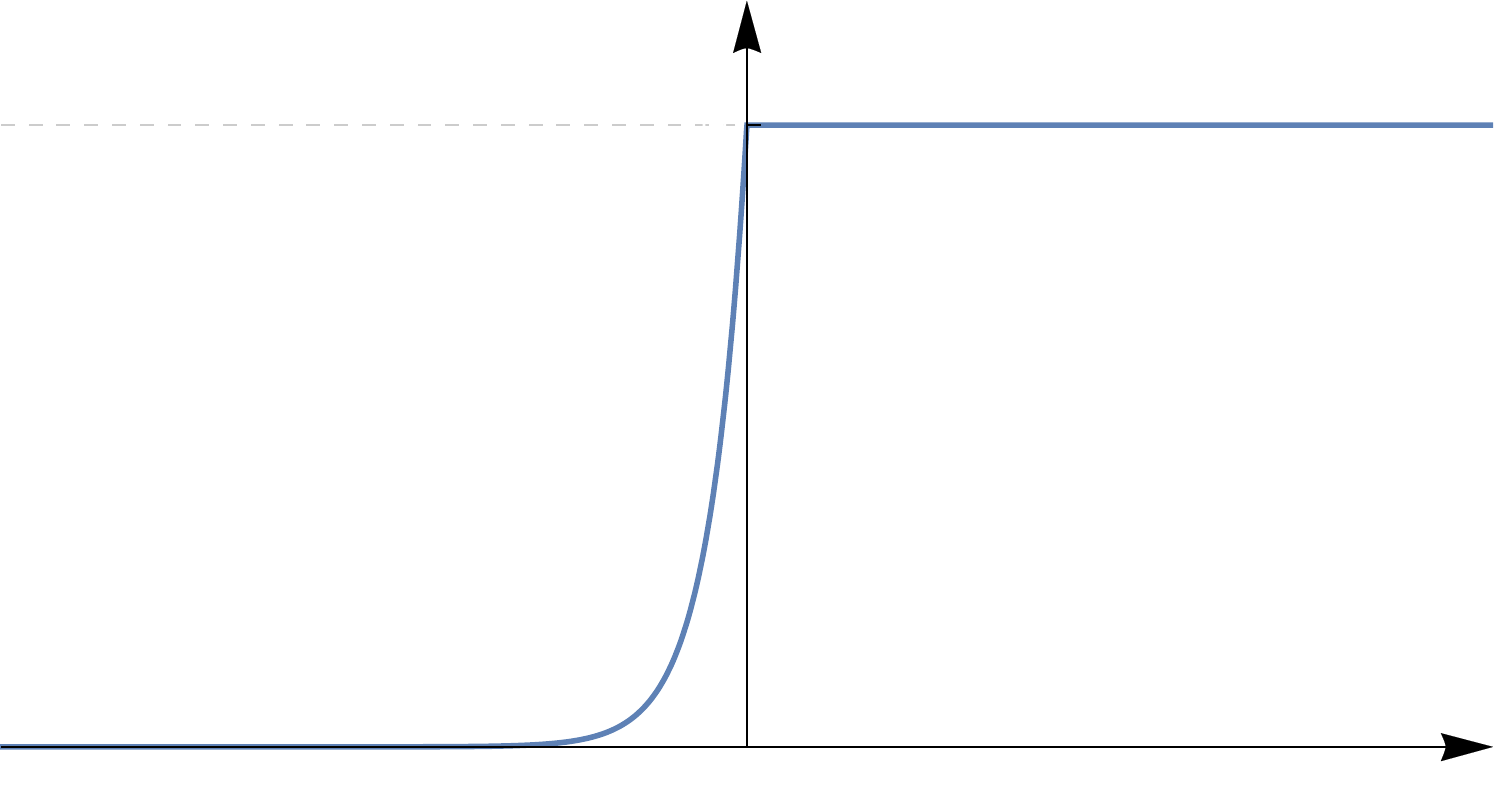}
			\put (43,46) {\footnotesize{$2M$}}
			\put (46,-1) {\footnotesize{$t=0$}}
			\put (48.5,56) {$\dot{\mathcal{C}}$}
			\put (102,2) {$t$}
		\end{overpic}
	\end{center}
	\caption{
		Growth rate~$\dot{\mathcal{C}}$ of holographic complexity as a function of tortoise time $t$, using the 
		CA prescription, for classical gravitational collapse. Following an exponential onset, holographic complexity 
		grows linearly with time.
		}\label{fig:ca_classical}
\end{figure}
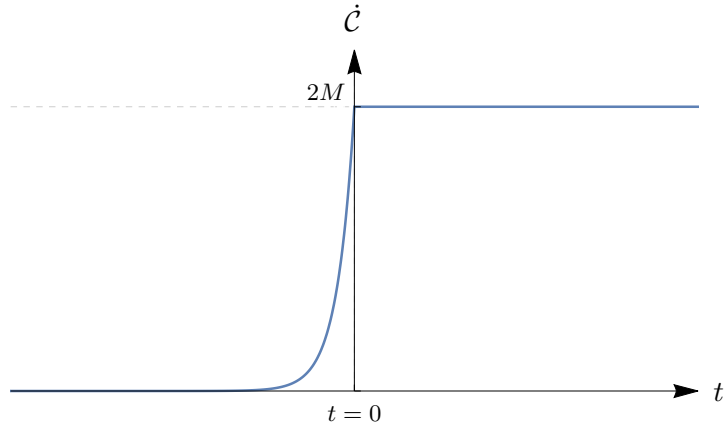

Our findings, summarized in Figure~\ref{fig:ca_classical}, are consistent with the expectation that the complexity 
of the quantum state corresponding to a black hole should grow with a rate proportional to the black hole entropy 
times its temperature.

\subsection{Eternal Black Hole}
For completeness we should mention that our prescription also works for the classical eternal black hole for late time. Its solution in terms of the dilaton, in Kruskal coordinates, is given by 
\begin{equation}\label{eq:cl_eternal_bh}
	e^{-2\phi} = e^{-2\rho} = M-x^{+} x^{-}
	\, ,
\end{equation}
see e.g.~\cite{Thorlacius:1994ip}.
The black and white hole singularities are located on the curves defined by $M = x_S^{+} x_S^{-}$.

\begin{figure}[h]
	\vspace{0.5cm}
	\begin{center}
		\begin{overpic}[width=.45\textwidth]{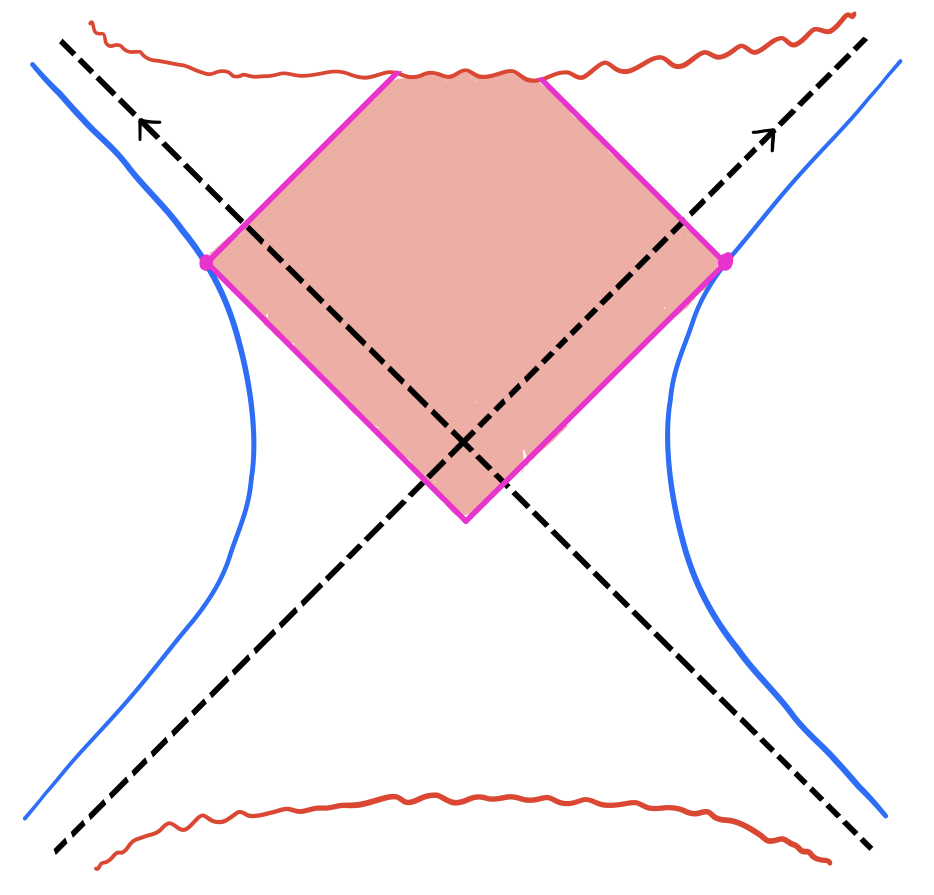}
			\put (20,82) {\footnotesize{$x^{-}$}}
			\put (75,82) {\footnotesize{$x^{+}$}}
			\put (43.5,64) {\footnotesize{Patch}}
			\put (43,68) {\footnotesize{WdW}}
			\put (26,2) {\footnotesize{White Hole Singularity}}
			\put (26,92) {\footnotesize{Black Hole Singularity}}
			\put (16,66) {\footnotesize{$t_{L}$}}
			\put (80,66) {\footnotesize{$t_{R}$}}
		\end{overpic}
	\vspace{-0.25cm}
	\end{center}
\caption{
Kruskal diagram of eternal black hole. A symmetric WdW patch is presented. Color coding agrees with previous figures.
}\label{fig:kruskal_eternal_bh}
\end{figure}
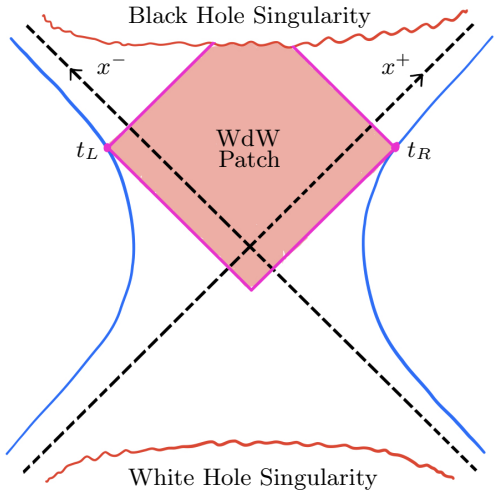

A Kruskal diagram including the WdW patch for late times is given in Figure~\ref{fig:kruskal_eternal_bh}. As is usual for the CA prescription in the context of two-sided black holes, we have to provide a second anchor point on a `left' anchor curve, see e.g. \cite{Carmi:2016wjl,Carmi:2017jqz}. The result for complexity growth will then be a function of $t_R-t_L$ where $t_R$ ($t_L$) are the tortoise times w.r.t.~the right (left) side associated to the respective anchor point position, see Figure~\ref{fig:kruskal_eternal_bh}. For simplicity, we take the two anchor points to move symmetrically as time progresses, {\it i.e.}~$t_L = -t_R$, so that the result only depends on $t = t_R$. In this case, we expect the complexity growth to have twice the contribution of a one-sided black hole.

The variation of the Kruskal area~$\mathcal{A}$ is easily performed and indeed provides the expected holographic complexity growth
\begin{equation}\label{eq:ca_semicl_eternal_bh}
	\dot{\mathcal{C}} = 4 M
	\, ,
\end{equation}
in the late time limit.

\section{Semi-Classical Black Hole Complexity}\label{section:blackholes_evap}
\subsection{Evaporating Black Hole}

Again, we study an incoming leftmoving shockwave pulse of energy $M$ at $x^{+}=x^{+}_{0}$ of the form
\begin{equation}
	T^{f}_{++}
	=
	\frac{M}{x_{0}^{+}}\delta(x^{+}-x^{+}_{0})
	\,.
\end{equation}
The semi-classical collapse solution, in terms of the field $\Omega$ defined in~\eqref{eq:def_omega_chi}, using Kruskal coordinates, is given by \cite{Russo:1992ax}
\begin{equation}\label{eq:q_coll_sol}
	\Omega(x^{+},x^{-}) =
		-x^{+}x^{-}+\left(x^{+}_{0}-x^{+}\right)\frac{M}{x^{+}_0}\Theta(x^{+}-x^{+}_{0})
		-\frac{\kappa}{4}\ln\left(-x^{+}x^{-}\right)
	\,,
\end{equation}
which in turn determines the dilaton~$\phi$ and the metric via its conformal factor~$\rho=\phi$.
The field~$\eta$ (see \eqref{eq:rst_bulk_rewrite}), needed for the holographic complexity computation, is related to the conformal anomaly of the energy momentum tensor~$T^f_{\mu \nu}$, $\langle T^f_{\mu}{^\mu} \rangle=\frac{\kappa}{2}R$, which fixes the form of the energy momentum tensor in light cone coordinates as~\cite{Christensen:1977jc,Callan:1992rs}
\begin{equation}
	\langle T^f_{\pm \pm} \rangle = -\kappa
	\left[
		\partial_{\pm} \rho \partial_{\pm} \rho - \partial_{\pm}^2 \rho + t_\pm(y^{\pm})
	\right]
	,
\end{equation}
where
\begin{equation}\label{eq:t_xi_rel}
	t_\pm =
		\frac{1}{4}\left(
			\partial_{\pm} \eta \partial_{\pm} \eta
			+ 2 \partial_{\pm}^2 \eta
		\right)
\end{equation}
is determined by boundary conditions imposed at past null infinity~$\mathcal{I}^{-}$, stating that there should be no outgoing energy flux at~$\mathcal{I}^{-}$. In this form, the energy momentum `tensor', including its quantum corrections, is actually not a tensor anymore. This is related to the fact, that the quantum corrections depend on a choice of vacuum, which makes reference to a specific coordinate system. For the case at hand, this is the coordinate system where the metric is manifestly Minkowskian near~$\mathcal{I}^{-}$.

The form of $t_{\pm}$ in Kruskal coordinates can be determined \cite{Callan:1992rs} to be $t_{\pm}(x^{\pm}) = -1/(2x^{\pm})^{2}$. 
The field~$\eta$ is fixed by~\eqref{eq:t_xi_rel} and reads \cite{Hayward:1994dw}
\begin{equation}\label{eq:xi_sol}
	\eta(x^+, x^-) = \log (-x^{+} x^{-})
	\, ,
\end{equation}
which is needed for the evaluation of holographic complexity, see equation~\eqref{eq:total_onshell_action}. Since the field~$\eta$ determines the functions $t_\pm$ (at least in Kruskal coordinates), it can be viewed as encoding the boundary conditions of the energy momentum tensor~$\langle T^f \rangle$. This is in line with the fact that the field~$\eta$ only appears within a total derivative term in the action~\eqref{eq:rst_bulk_rewrite}.

The form of $\langle T^f_{++} \rangle$ implies, that an asymptotic observer near future null infinity~$\mathcal{I}^+$ will 
see a non-vanishing outgoing matter energy flux, {\it i.e.} Hawking radiation, which turns on with an exponential onset
as the black hole is formed and turns off when the mass of the black hole has been depleted. The endpoint of Hawking 
emission is abrupt in the RST model and even requires a small adjustment in the form of a negative energy shock wave 
emanating from the black hole endpoint.\footnote{See {\it e.g.} \cite{Thorlacius:1994ip} for more details.} This reflects a 
breakdown of the semi-classical description when the remaining black hole mass approaches the Planck scale and serves
as a reminder that our semi-classical calculation of holographic complexity will be subject to similar limitations as the
black hole mass is depleted.

\begin{figure}[h]
	\vspace{0.5cm}
	\begin{center}
		\begin{overpic}[width=.4\textwidth]{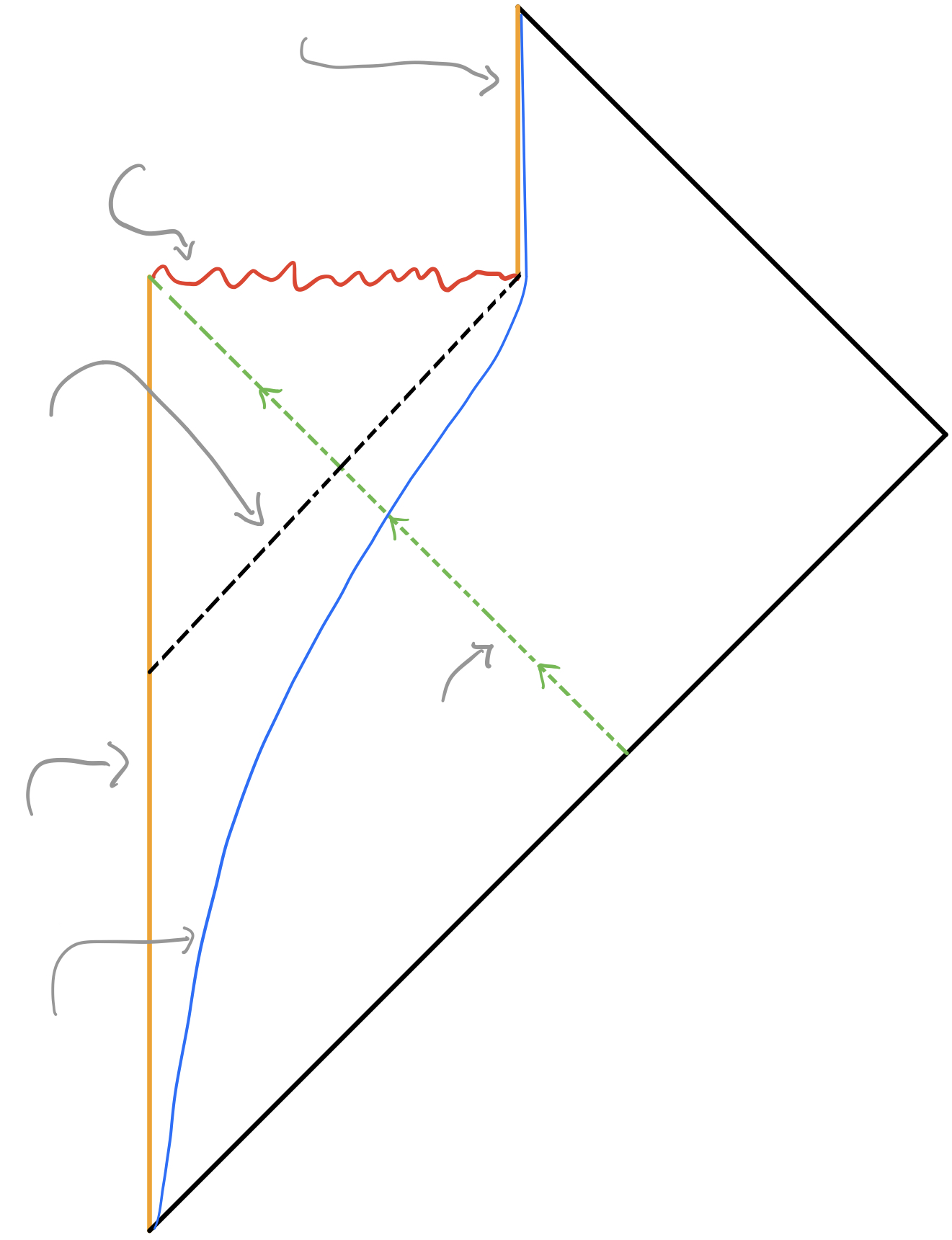}
			\put (-6,30) {\footnotesize{$\Omega=\Omega_{\text{crit}}$}}
			\put (19,97.5) {\footnotesize{$\Omega=\Omega_{\text{crit}}$}}
			\put (-6,14) {\footnotesize{Stretched}}
			\put (-4.5,10.5) {\footnotesize{Horizon}}
			\put (12.5,86) {\footnotesize{Black Hole}}
			\put (12.5,82.5) {\footnotesize{Singularity}}
			\put (-1,63) {\footnotesize{Event}}
			\put (-3,59) {\footnotesize{Horizon}}
			\put (31,40) {\footnotesize{Matter}}
			\put (31,36.5) {\footnotesize{Fields}}
			\put (38,100) {\footnotesize{$i^{+}$}}
			\put (78,64) {\footnotesize{$i^{0}$}}
			\put (11,-3) {\footnotesize{$i^{-}$}}
			\put (48,33) {\footnotesize{$\mathcal{I}^{-}$}}
			\put (59,84) {\footnotesize{$\mathcal{I}^{+}$}}
		\end{overpic}
		\hspace{.5cm}
		\begin{overpic}[width=.35\textwidth]{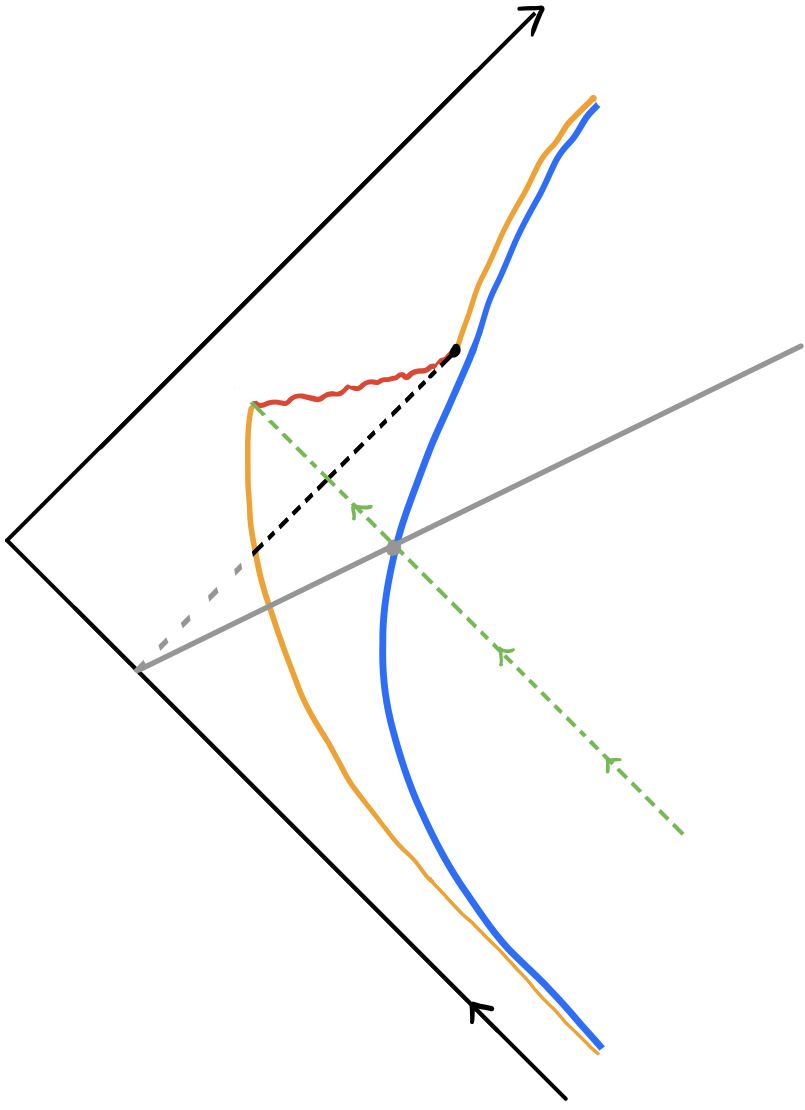}
			\put (20,23) {\footnotesize{$x^{-}$}}
			\put (18,74) {\footnotesize{$x^{+}$}}
			\put (54,14) {\footnotesize{Horizon}}
			\put (52,18) {\footnotesize{Stretched}}
			\put (50.5,39) {\rotatebox{-45}{\footnotesize{$x^{+}\!=x^{+}_{0}$}}}
			\put (61,66) {\rotatebox{25}{\footnotesize{$t=0$}}}	
		\end{overpic}
	\end{center}
\caption{
Left panel: Penrose diagram depicting the life cycle of evaporating black hole formed by collapse. 
Right panel: The corresponding Kruskal diagram with the same color coding. The gray line denotes 
a curve of equal tortoise time $t$.
}\label{fig:evaporation}
\end{figure}
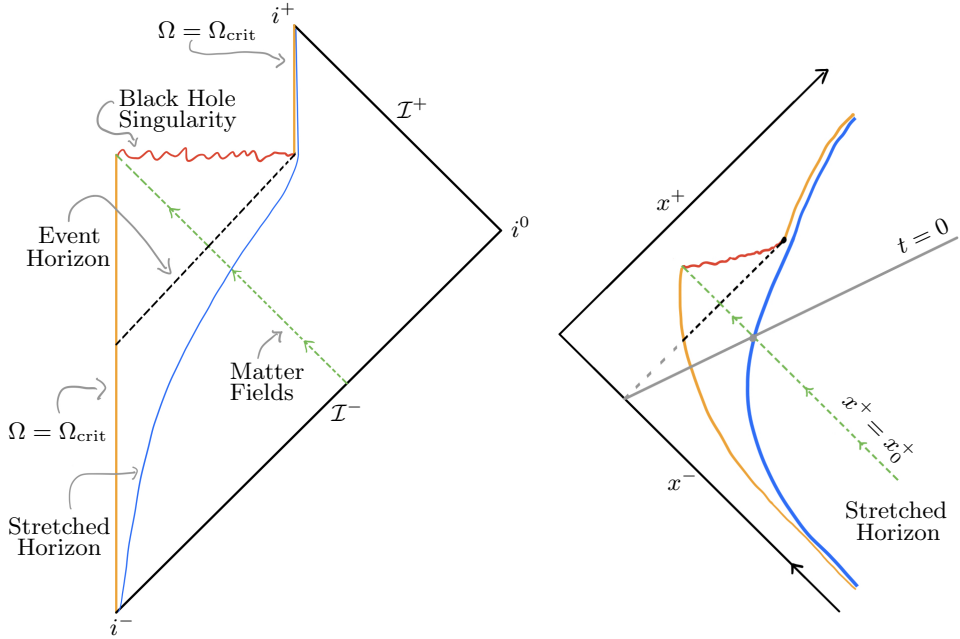

Altogether, the solution~\eqref{eq:q_coll_sol} together with~\eqref{eq:xi_sol}, describes flat spacetime for $x^{+} \leq x^{+}_0$ and an evaporating black hole for $x^{+} > x^{+}_0$ with outgoing Hawking radiation towards future infinity~$\mathcal{I}^{+}$, see Figure~\ref{fig:evaporation}.
The location of the black hole singularity is determined by the curve $(x^{+}_S,x^{-}_S)$ which satisfies $\Omega(x^{+}_S,x^{-}_S)=\Omega_\text{crit}=\frac{\kappa}{4}(1-\ln\frac{\kappa}{4})$ for $x^{+} > x^{+}_0$. A useful parametrization of this curve, which we employ at a later stage, is given by
\begin{equation}\label{eq:sing_parametrization}
	\begin{pmatrix}
		x^{+}_{S}(u) \\
		x^{-}_{S}(u)
		\end{pmatrix}
		=
	\begin{pmatrix}
		x^{+}_{0}\left(\frac{\kappa}{4M}\left(e^{\frac{4M}{\kappa}u}-1\right)-u+1\right) \\
		-\frac{\kappa}{4x^{+}_{0}}\frac{e^{\frac{4M}{\kappa}u}}{\frac{\kappa}{4M}\left(e^{\frac{4M}{\kappa}u}-1\right)-u+1}
	\end{pmatrix}
	\,,
\end{equation}
where the range of the parameter $u$ is the interval $(0,1)$. The point $(x^{+}_S(0),x^{-}_S(0))$ represents the formation
of the black hole singularity, while $(x^{+}_S(1),x^{-}_S(1))$ describes the point where the black hole has entirely evaporated. 
This parametrization has the convenient property
\begin{equation}
	- x^{+}_S(u) \, x^{-}_S(u) = \frac{\kappa}{4} e^{\frac{4M}{\kappa} u}
	\, .
\end{equation}
The curve $\Omega(x^{+}_B,x^{-}_B)=\Omega_\text{crit}=\frac{\kappa}{4}(1-\ln\frac{\kappa}{4})$ for $x^{+} < x^{+}_0$ 
defines the boundary of physical spacetime before the matter shockwave arrives, and is given by
\begin{equation}\label{eq:boundary_def_semicl}
	- x^{+}_B x^{-}_B = \frac{\kappa}{4}
	\, ,
\end{equation}
in Kruskal coordinates.

\begin{figure}[h]
	\begin{center}
		\begin{overpic}[width=.35\textwidth]{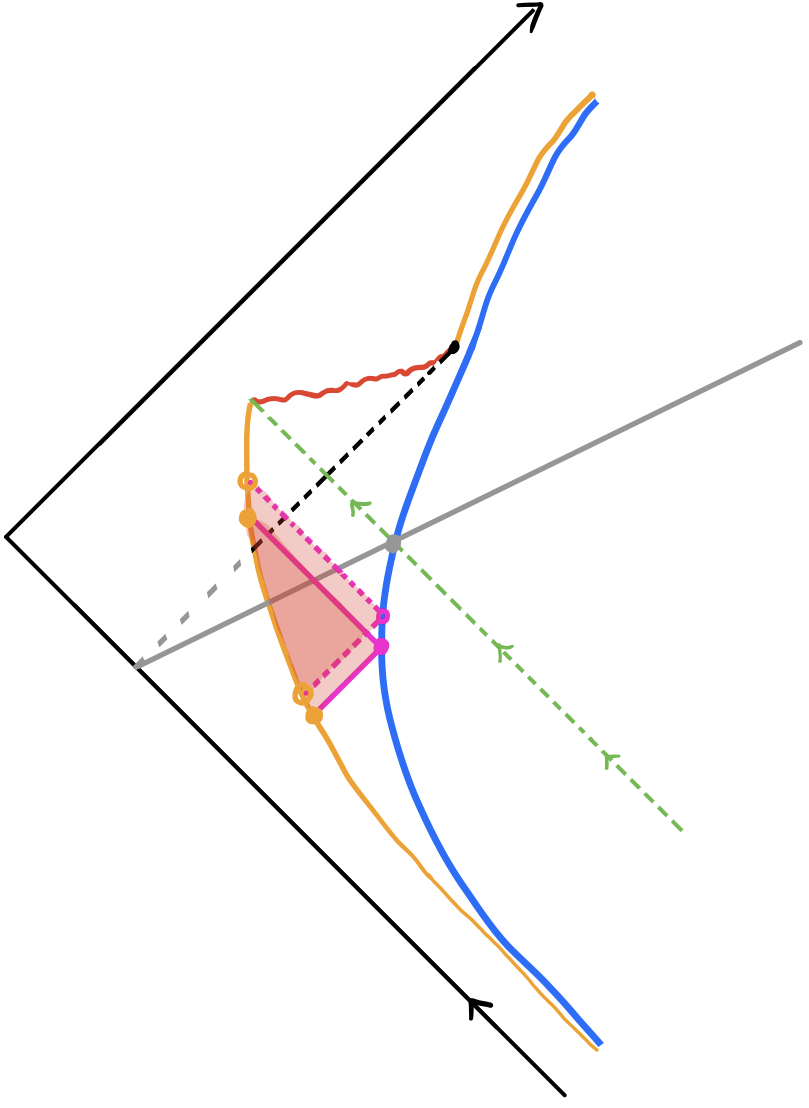}
			\put (35,8) {\footnotesize{$x^{-}$}}
			\put (30,87) {\footnotesize{$x^{+}$}}
			\put (48,14) {\footnotesize{Horizon}}
			\put (45,18) {\footnotesize{Stretched}}
%
%
%
			\put (51.5,37) {\rotatebox{-45}{\footnotesize{$x^{+}\!=x^{+}_{0}$}}}
			\put (61,66) {\rotatebox{25}{\footnotesize{$t=0$}}}	
%
		\end{overpic}
		\hspace{1.5cm}
		\begin{overpic}[width=.35\textwidth]{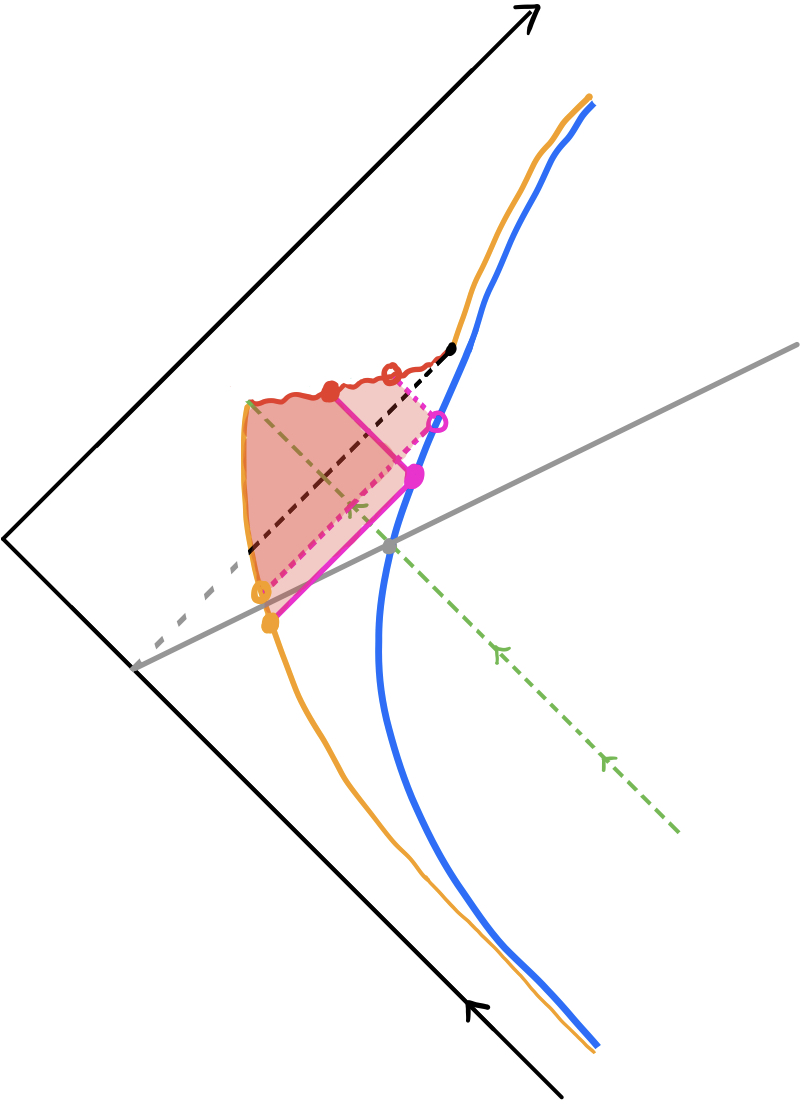}
			\put (11.5,32) {\footnotesize{$x^{-}$}}
			\put (20,76) {\footnotesize{$x^{+}$}}
			\put (48,14) {\footnotesize{Horizon}}
			\put (45,18) {\footnotesize{Stretched}}
			
%
			
			\put (51.5,37) {\rotatebox{-45}{\footnotesize{$x^{+}\!=x^{+}_{0}$}}}
			\put (61,66) {\rotatebox{25}{\footnotesize{$t=0$}}}	
%
		\end{overpic}
	\end{center}
\caption{
Kruskal diagrams of evolution of a WdW patch of an evaporating black hole. Color coding coincides with Figure \ref{fig:evaporation}. 
}\label{fig:evaporation_wdw}
\end{figure}
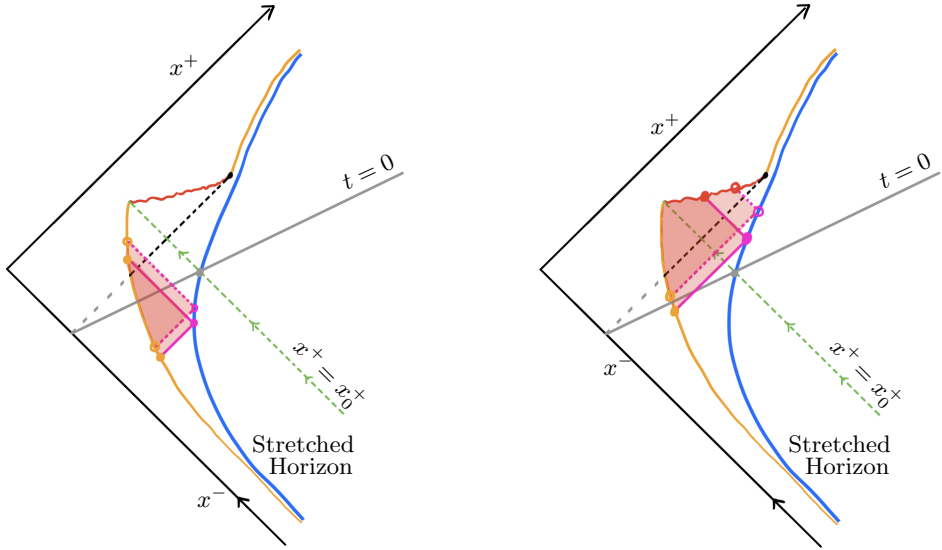

As in the classical theory, we take the anchor curve for our WdW patch to be the stretched horizon of the black hole,
defined as a membrane outside the black hole event horizon, with an area of order $1$, in Planck units, larger than 
the area of the black hole event horizon. For technical simplicity, we follow~\cite{Susskind:1993if} and take the 
stretched horizon of an RST black hole formed by shockwave collapse to coincide with the apparent horizon during
the period of evaporation. 
This determines the anchor curve as 
\begin{equation}\label{eq:def_semicl_sh}
	-x^{+}_A \left( x^{-}_A + \frac{M}{x^{+}_0} \right)
	= \frac{\kappa}{4}
	\, ,
\end{equation}
as usual, in Kruskal coordinates.
With this convention, the area of the stretched horizon vanishes at the evaporation end point, whereas it should
strictly speaking be $1$ in Planck units there. However, the error is negligible as long as the black hole remains 
large compared to the Planck scale, {\it i.e.} whenever the semi-classical approximation can be relied on in the first place.
With these ingredients, it is now possible to define the WdW patch in a similar fashion as in the classical case, 
see Figure~\ref{fig:evaporation_wdw}. 

The on-shell action~\eqref{eq:total_onshell_action}, together with the field~$\eta$ given by~\eqref{eq:xi_sol}, can be formulated as 
\begin{equation}
	\mathcal{C} = 2 \mathcal{A} + \frac{\kappa}{2} \mathcal{B}
	:= \int_\text{WdW} dx^{+} dx^{-}
	+ \frac{\kappa}{2} \int_\text{WdW} \frac{dx^+}{x^+} \frac{dx^-}{x^-}
	\, .
\end{equation}
In addition to the `flat' reference metric area~$\mathcal{A}$ in Kruskal coordinates, the semi-classical holographic complexity 
acquires a correction~$\frac{\kappa}{2}\mathcal{B}$. For future evaluation purposes, we note that the correction term can also 
be given an `area' interpretation, by changing from Kruskal coordinates to their logarithm,
\begin{equation}
	\begin{aligned}
	\sigma^+ & = \log (x^+) \\
	\sigma^- & = \log (-x^-)
	\,.
	\end{aligned}
\end{equation}
To evaluate holographic complexity we again consider two cases: the WdW patch anchored in the region before the shock wave 
arrives, $x^{+} < x^{+}_0$, and after, $x^{+} > x^{+}_0$.

\paragraph{Before incoming shockwave}
The evaluation of the change of area~$\mathcal{A}$ is completely analogous to the classical case, see Figure~\ref{fig:evaporation_wdw}. We have 
\begin{equation}
	d \mathcal{A}
	= \left( x^{-}_{B}(x^{+}_A) - x^{-}_A \right) d x^{+}_A
	- \left( x^{+}_A - x^{+}_B ( x^{-}_A ) \right)  d x^{-}_A
	\, ,
\end{equation}
which can be evaluated,
making use of~\eqref{eq:boundary_def_semicl} and \eqref{eq:def_semicl_sh},
 to give
\begin{equation}
	d \mathcal{A}
	= \left( M e^t - \frac{\kappa}{4}  + \mathcal{O}\left(M^{-1}\right) \right) dt
	\, .
\end{equation}
The result deviates from the classical calculation by a constant contribution proportional to~$\kappa$.

Similarly,
\begin{equation}
	d \mathcal{B}
	= \log \left( \frac{x^{-}_{B}(x^{+}_A)}{x^{-}_A} \right) d \log(x^{+}_A)
	- \log \left( \frac{x^{+}_A}{x^{+}_B ( x^{-}_A )}  \right) d \log(x^{-}_A)
	\, .
\end{equation}
The result is exponentially suppressed as $t\to -\infty$, but gives a non-negligible contribution for times near the black hole creation ($t=0$) of the form
\begin{equation}
	\frac{\kappa}{2} d \mathcal{B}
	\approx
	- \frac{\kappa}{2} \left( \log \left(\frac{4 M}{\kappa }\right)+t \right) dt
	\, .
\end{equation}

\paragraph{After incoming shockwave}
Replacing the spacetime boundary by the singularity curve in the region $x^{+} > x^{+}_0$ gives the correct change of Kruskal area,
\begin{equation}
	d \mathcal{A}
	= \left( x^{-}_{S}(x^{+}_A) - x^{-}_A \right) d x^{+}_A
	- \left( x^{+}_A - x^{+}_B ( x^{-}_A ) \right)  d x^{-}_A
	\, .
\end{equation}
In addition to \eqref{eq:boundary_def_semicl} and \eqref{eq:def_semicl_sh}, using the parametrization~\eqref{eq:sing_parametrization}, we can express the result as
\begin{equation}
	d \mathcal{A}
	= M(1-u) \, dt - \frac{\kappa}{4} dt + \mathcal{O}(e^{-\frac{4M}{\kappa}u})
	\, ,
\end{equation}
where corrections are suppressed after a scrambling time $t_S = \log(\frac{4M}{\kappa})$.
The parameter~$u$ is related to time $t$ via
\begin{equation}
	\frac{4M}{\kappa} \left( e^{t} -1 \right) = e^{\frac{4M}{\kappa}u} - \frac{4M}{\kappa} u -1
	\, ,
\end{equation}
which, for times after the scrambling time~$t_S$, can be expressed as
\begin{equation}
	u(t \gtrapprox t_S)
	\approx
	\frac{\kappa}{4M}\left(t + \log\left(\frac{4M}{\kappa}\right)\right)
	\, .
\end{equation}
Notably, the growth of the Kruskal area~$\mathcal{A}$ with time $t$ is linear after the scrambling time. Moreover, the result for $\frac{d\mathcal{A}}{dt}$ is continuous at the black hole creation time $t=0$.

That leaves us with the evaluation of the logarithmic area~$\mathcal{B}$,
\begin{equation}
	d \mathcal{B}
	= \log \left( \frac{x^{-}_{S}(x^{+}_A)}{x^{-}_A} \right) d \log(x^{+}_A)
	- \log \left( \frac{x^{+}_A}{x^{+}_B ( x^{-}_A )}  \right) d \log(x^{-}_A)
	\, .
\end{equation}
The result is suppressed after the scrambling time~$t_S$, but contributes near the black hole creation time $t=0$,
\begin{equation}
	\frac{\kappa}{2} d \mathcal{B}
	\approx
	- \frac{\kappa}{2} \log \left(\frac{4 M}{\kappa }\right) dt
	\,
\end{equation}
which shows that also $\frac{d\mathcal{B}}{dt}$ is continuous at $t=0$.

It is interesting to observe, that the contribution of $\mathcal{B}$ before the scrambling time~$t_S$ conspires with 
the non-linear contribution of $\mathcal{A}$ before the scrambling time to provide linear growth up to corrections 
of order $M^{-1}$ even before the scrambling time. The final exact result for the complexity growth rate after the 
black hole creation is given by
\begin{equation}\label{eq:final_res}
	\begin{aligned}
	\dot{\mathcal{C}}(t) & =
	2M\left(1-\frac{t+1+\log\left(\frac{4M}{\kappa}+e^{-t}\right)}{\frac{4M}{\kappa}+e^{-t}}\right)
	\\
	& =
	2 M - \frac{\kappa}{2} \left(
	\log \left(\frac{4 M}{\kappa }\right) + 1 + t 
	+ \mathcal{O} \left( \frac{\kappa}{M}e^{-t} \right)
	\right)
	\, .
	\end{aligned}
\end{equation}
This result holds until the black hole has fully evaporated, but, as stated above, it becomes unreliable when
the remaining mass is of order the Planck scale.

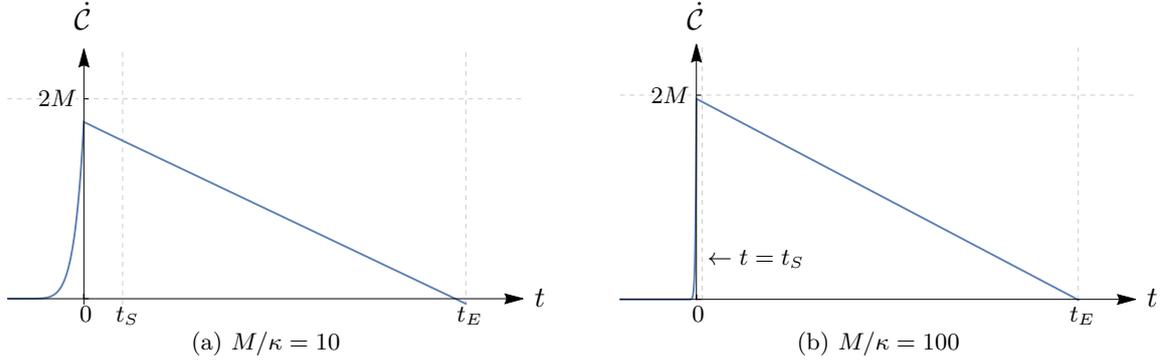
\begin{figure}
	\begin{center}
		\subfloat[$M/\kappa=10$]{
			\begin{overpic}[width=.45\textwidth]{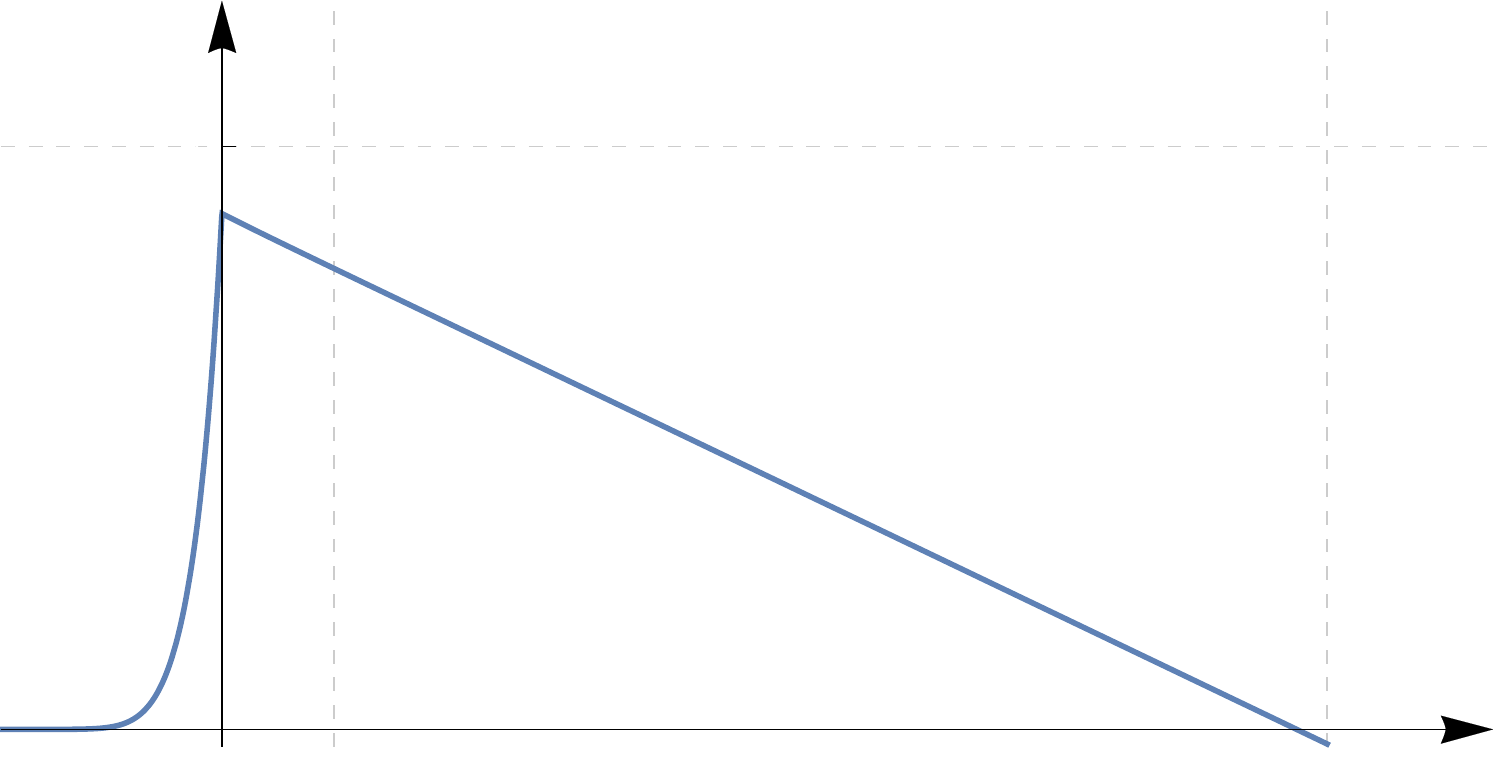}
				\put (6,41) {\footnotesize{$2M$}}
				\put (14,-1) {\footnotesize{$0$}}
				\put (21,-1) {\footnotesize{$t_S$}}
				\put (87,-1) {\footnotesize{$t_E$}}
				\put (13,56) {$\dot{\mathcal{C}}$}
				\put (102,2) {$t$}
			\end{overpic}
		}
		\hfill
		\subfloat[$M/\kappa=100$]{
			\begin{overpic}[width=.45\textwidth]{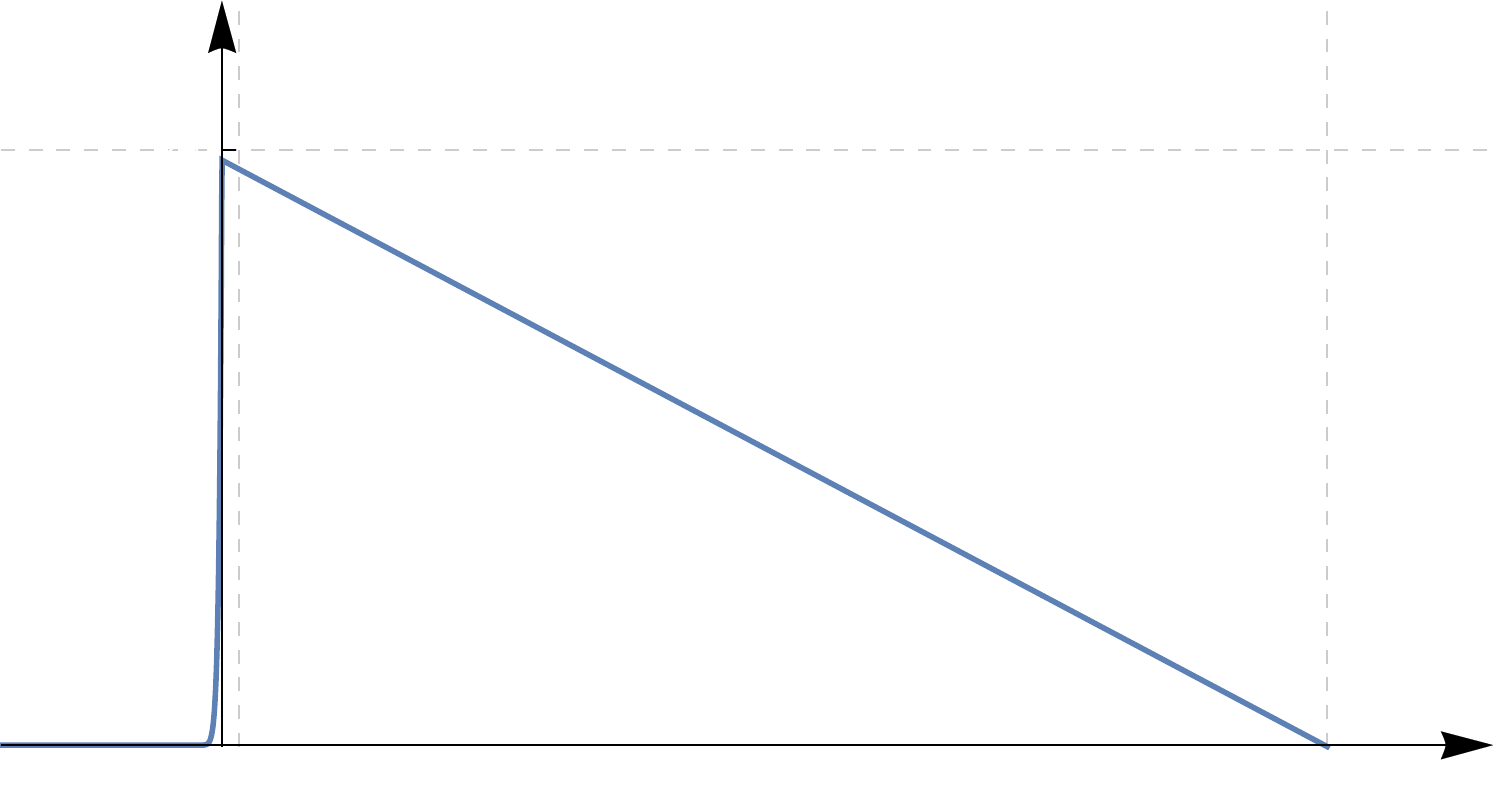}
				\put (6,41.5) {\footnotesize{$2M$}}
				\put (14,-1) {\footnotesize{$0$}}
				\put (17,10) {\footnotesize{$\leftarrow t=t_S$}}
				\put (87,-1) {\footnotesize{$t_E$}}
				\put (13,56) {$\dot{\mathcal{C}}$}
				\put (102,2) {$t$}
			\end{overpic}
		}
	\end{center}
	\caption{
		The growth rate~$\dot{\mathcal{C}}$ of holographic complexity as a function of tortoise time $t$, using the CA prescription, 
		for  evaporating black holes of different initial mass. The exponential onset at the creation time of the black hole is followed 
		by a linear falloff period until the black hole has evaporated.
		}\label{fig:ca_semi-classical}
\end{figure}

The total result is plotted in Figure~\ref{fig:ca_semi-classical}. The plot confirms the continuity and  
linear falloff of the growth rate of holographic complexity~$\dot{\mathcal{C}}$.
A short time before the lifetime~$t_E$ of the black hole has been reached, the value of the complexity growth rate hits zero and subsequently becomes slightly negative. This is potentially problematic, since there is no reason to believe, that the complexity growth rate of an evaporating black hole should ever be negative. However, at that time, the mass of the black hole has already attained the Planck scale, and, as stated above, our result is not trustworthy anymore.

\subsection{Eternal Black Hole}
We can also study a semi-classical eternal black hole by including a heat bath at spatial infinity, with a temperature
equal to the Hawking temperature of the black hole. The heat bath provides a steady 
incoming energy flux which matches the outgoing flux from the radiating black hole. 
The solution, in Kruskal coordinates, is given by \cite{Thorlacius:1994ip}
\begin{equation}\label{eq:semicl_eternal_bh}
	\Omega (x^{+}, x^{-})= -x^{+}x^{-} + M + \frac{\kappa}{4} - \frac{\kappa}{4} \log \left( \frac{\kappa}{4} \right)
	\, .
\end{equation}
The spacetime curvature is singular where $\Omega = \Omega_\text{crit}$, {\it i.e.} on curves satisfying
\begin{equation}\label{eq:semicl_eternal_bh_sing}
	x^{+}_S x^{-}_S = M
	\, ,
\end{equation}
describing a black hole and white hole singularity. These are the same curves as for the 
singularities of the classical eternal black hole described by~\eqref{eq:cl_eternal_bh}.
The Kruskal diagram for a semi-classical eternal black hole solution is thus 
identical to that of a classical eternal black hole, shown in Figure~\ref{fig:kruskal_eternal_bh}, 
but the physics described by the semi-classical solution is somewhat different. 
In contrast to all other solutions considered in this work, the parameter~$M$ in~\eqref{eq:semicl_eternal_bh} is \emph{not} 
proportional to the ADM mass of the black hole. Since the semi-classical solution describes a black hole in equilibrium 
with a heat bath at infinity there is non-vanishing 
energy density in the asymptotic region and the ADM mass diverges.\footnote{The infinite train of radiation does not lead 
to a catastrophic back-reaction on the geometry because the gravitational coupling, governed by $e^\phi$, goes rapidly to 
zero asymptotically.} The parameter~$M$ is characteristic of the black hole size and therefore we will continue to 
refer to it as the `mass' of the black hole. 

Since the semi-classical Kruskal diagram is unchanged compared to the diagram of a classical black hole, 
and the Kruskal area~$\mathcal{A}$ is only sensitive to the location of the singularity and not the detailed form 
of the dilaton field, it agrees with the classical calculation,
\begin{equation}
	d\mathcal{A} = 2M dt
	\, ,
\end{equation}
for late times.
Further, it follows from~\eqref{eq:semicl_eternal_bh} that $t_{\pm}=0$ in Kruskal coordinates and then 
equation \eqref{eq:t_xi_rel} immediately implies that $\eta=0$. It follows that the semi-classical correction~$\mathcal{B}$ 
vanishes. 

We conclude that the complexity growth of the semi-classical eternal black hole for late times agrees with the 
classical result~\eqref{eq:ca_semicl_eternal_bh},
\begin{equation}\label{eternalgrowth}
	\dot{\mathcal{C}} = 4 M
	\, ,
\end{equation}
and does not receive semi-classical corrections.

\section{Discussion and outlook}\label{section:summary}
In this paper we have investigated the holographic complexity of evaporating black holes in a
toy model where the semi-classical geometry is known explicitly. The CA proposal for black hole
complexity can be adapted to this model and we have obtained analytic expressions for the 
increase in complexity over the lifetime of a semi-classical black hole. This extends our 
previous work in~\cite{Schneiderbauer:2019anh} where we numerically evaluated the semi-classical 
complexity using a CV prescription for the same model. The analytic CA results presented here 
provide a much more detailed picture of how complexity evolves as the black hole evaporates
compared to the previous numerical CV evaluation. For parameter values where a meaningful 
comparison can be carried out, the two approaches are in good agreement, starting from a scrambling
time after the black hole is formed and for the remainder of the black hole lifetime. 

In order to ensure a well-posed variational principle for the action on a Wheeler-DeWitt patch, it is 
necessary to include appropriate boundary terms in the action. These boundary terms are not unique, 
something that is true for CA in general, but for a range of black holes in classical Einstein gravity
the ambiguity does not affect the late time rate of increase of complexity \cite{Lehner:2016vdi}. 
In the context of semi-classical black holes, the finite black hole lifetime limits the ability
to take a late time limit and the ambiguity involving boundary terms needs to be addressed 
in order to have a well-defined CA prescription. This can be achieved in a natural way in the
RST model by extending a symmetry of the original semi-classical bulk theory to the boundary
terms as well. The final analytic result for complexity growth rate during the evaporation process, 
presented in formula~\eqref{eq:final_res}, has several interesting features.

First, it confirms linear falloff of $\dot{\mathcal{C}}$ with time after the scrambling time~$t_S$, already 
observed (numerically) in~\cite{Schneiderbauer:2019anh} using a CV prescription. In fact, up to
small corrections, equation \eqref{eq:final_res} exhibits linear behaviour already from $t=0$, the time 
of black hole formation, in contrast to CV where the numerics indicates an initial adjustment period of 
order the scrambling time.
The linear falloff is important, as it captures the time evolution of the entropy of the evaporating black hole. 
Classical black hole entropy is given by $S_0=2M$ in this model and at the semi-classical level the
black hole radiates mass at a constant rate~$\kappa/4$, so that
\begin{equation}\label{massevolution}
	S_0(t) 
	= 
	2M(t)
	=
	2M - \frac{\kappa}{2} t \,.
\end{equation}
The Hawking temperature is independent of black hole mass in this model so the relation 
\begin{equation}\label{growthrelation}
	\dot{\mathcal{C}}(t) \propto S(t)~T \,,
\end{equation}
is seen to hold at leading order in a $\kappa/M$ expansion for large initial black hole mass.

Second, the subleading logarithmic term in the rate of complexity increase in~\eqref{eq:final_res}
can also be given an interpretation in terms of entropy. 
In~\cite{Solodukhin:1995te,Myers1994}, it was shown that the leading order quantum-corrected 
entropy for a semi-classical black hole in equilibrium with a thermal heat bath, is given by 
\begin{equation}\label{eq:entropy}
	S = 2e^{-2\phi_h} + \frac{\kappa}{2} \phi_h - \frac{\kappa}{2}+\frac{\kappa}{4}\log{\frac{\kappa}{4}}
	\, ,
\end{equation}
where $\phi_h$ is the value of the dilaton field at the horizon. When evaluated for a dynamical solution of the 
RST model describing a black hole formed by an incoming shock wave, this gives
\begin{equation}
	S = 2 M - \frac{\kappa}{2} \log \left( \frac{4M}{\kappa} \right)-\frac{\kappa}{2}
	\, ,
\end{equation}
immediately after the black hole is formed and zero at the evaporation endpoint.
Comparing to~\eqref{eq:final_res} shows that the rate of complexity growth at the 
onset of black hole evaporation is consistent with the relation~\eqref{growthrelation}, 
even including the subleading logarithmic term.\footnote{Due to the slow evolution of the 
logarithm, this remains true for the bulk of the black hole lifetime.} 
If we instead evaluate the entropy formula~\eqref{eq:entropy} for the static 
solution \eqref{eq:semicl_eternal_bh}, describing an eternal black hole in equilibrium
with a heat bath, we find that the entropy takes its classical value,
\begin{equation}
S=2M \,.
\end{equation}
The cancellation of the semi-classical corrections can ultimately be traced to the 
back-reaction on the spacetime geometry due to the matched ingoing and outgoing
radiation flux \cite{Solodukhin:1995te}.
Interestingly, the corresponding cancellation also takes place in the rate of complexity
increase \eqref{eternalgrowth} for an eternal RST black hole and we once again find
that the relation \eqref{growthrelation} holds with $S(t)$ given by the semi-classical
entropy.

One may wonder how to interpret our formulas after the black hole has evaporated. 
One can still define a stretched horizon as the timelike curve where the transverse area 
is one unit larger than zero. This curve is very close to the boundary at $\phi=\phi_{\text{crit}}$ 
and a WdW patch anchored on it only covers a microscopic spacetime region.\footnote{The 
same is of course true for a WdW patch at very early times, long before the black hole is formed.} 
Both the Kruskal area $\mathcal A$ and the semi-classical correction $\mathcal B$ will have minuscule 
values, which do not change with time. This is consistent with zero growth of the
holographic complexity at late times, but having a vanishingly small WdW patch action 
at late times does not reflect the very large absolute complexity that was built up during
the evaporation of the black hole and is carried in the outgoing train of Hawking radiation.
For a classical black hole, the growth of the WdW patch action continues indefinitely and this
issue does not arise. An obvious way around this is to only use the action prescription to 
calculate the change in complexity and simply define the 
absolute holographic complexity as the integral of $\dot{\mathcal C}$ over time.

The constant Hawking temperature
of CGHS and RST black holes simplifies calculations but is rather unphysical. It would be 
interesting to study the charged version of the CGHS black holes, see 
{\it e.g.}~\cite{Frolov:2005ps,Frolov:2006is}, 
or the semi-classically corrected Jackiw-Teitelboim (JT) model, see {\it e.g.}~\cite{Almheiri2015}, 
since black holes in those two-dimensional models have varying temperature.

\subsection*{Acknowledgments}
It is a pleasure to thank Shira Chapman and Nick Poovuttikul for discussions.
This research is supported by the Icelandic Research Fund under grants 185371-051 and 195970-051, 
and by the University of Iceland Research Fund.

\appendix
\section{Stokes' theorem in two dimensions with null boundaries}\label{section:stoke_null}
Stokes' theorem in the context of Lorentzian manifolds is usually presented for manifolds with 
spacelike or timelike boundaries. For simplicity, we focus on a single smooth boundary component~$\partial M$. 
The result for piecewise smooth boundaries is obtained by summing over individual boundary components.
The theorem states that
\begin{equation}\label{stokes1}
	\int_{M}\sqrt{|g|}\nabla_{\mu}j^{\mu}=\int_{\partial M}\sqrt{|h|}n^{\mu}j_{\mu}+\ldots \,
\end{equation}
where $n^\mu$ is a inwards (outwards) pointing unit normal vector if $\partial M$ is spacelike (timelike) and the 
$\ldots$ indicates contributions from other boundary components. The integrals naturally involve the metric 
determinant $|g|$ and the determinant of the induced metric $|h|$ at the boundary~$\partial M$.

The expression on the right hand side of \eqref{stokes1} does not make sense for null boundaries, 
as the induced metric $h$ and the unit normal vector $n^\mu$ are degenerate in this case.
Since Stokes' theorem in its general form is a statement involving differential forms, it is
oblivious to a metric on a manifold.\footnote{The need for a metric only arises if one wants to 
integrate scalar functions over a manifold in a coordinate-invariant way.}
Thus Stokes' theorem also has to be valid for manifolds with null boundaries.
Our goal in this appendix is to find a simple expression to replace \eqref{stokes1} for null boundary components
in a two-dimensional context. This is easily achieved using a limiting procedure where the null boundary
curve is approximated by a family of either timelike or spacelike curves.

We work in conformal gauge,
\begin{equation}
	ds^2 = - e^{2\rho} dy^{+} dy^{-}
	\, ,
\end{equation}
with light-like coordinates $(y^+,y^-)$ and consider a null boundary component, $\partial M$, that we 
take to lie in the future of $M$ and described by a null curve of the form $y^- = y^-_0=\text{const}$.
In a region near the point $(y^+_0,y^-_0)$, the null curve is approached by a family of timelike curves
\begin{equation}\label{tcurve}
y^- = y^-_0 +\epsilon\, (y^+-y^+_0)\,,
\end{equation}
with $\epsilon>0$, in the limit $\epsilon\rightarrow 0$.

The outward-directed unit normal to the timelike curve in \eqref{tcurve} is given by
\begin{equation}
	n=e^{-\rho}\left(-\frac{1}{\sqrt{\epsilon}}\partial_{+}+\sqrt{\epsilon}\,\partial_{-}\right)\,,
\end{equation}
and the determinant of the induced metric evaluates to
\begin{equation}
	\sqrt{|h|}=e^{\rho}\sqrt{\epsilon}
	\, ,
\end{equation}
thus giving
\begin{equation}\label{eq:null_bdr_stokes_ex}
	\lim_{\epsilon\to0}\int_{\partial M}\sqrt{|h|}n^{\mu}j_{\mu}=-\int_{\partial M}dy^{+}j_{+}
	\, .
\end{equation}
We could also have approximated the null curve~$y^- = y^-_0$ by a family of spacelike curves and
considered the past directed normal vector. The resulting limit agrees with~\eqref{eq:null_bdr_stokes_ex}.

Furthermore, a similar procedure can be applied for null curves defined by $y^+ = \text{const}$ and 
for null boundaries that lie in the past of $M$. The general result, for the case of $M$ only having null 
boundaries defined by $y^\mp = \text{const}$, and not making reference to a particular coordinate system, 
can be written as 
\begin{equation}
	\int_{M}\sqrt{|g|}\nabla_{\mu}j^{\mu}
	=
	\sum_{\mathcal{N}\in\mathscr{N}}\sigma_{\mathcal{N}} \int_{\mathcal{N}} d\lambda \, k_\mathcal{N}^{\mu} j_{\mu}
\end{equation}
where the sum runs over all piecewise smooth null boundary components. The future-directed null vector $k^\mu$, tangential to the null boundary~$y^\mp = \text{const}$, is introduced, such that $\partial_\lambda = k^\mu \partial_\mu$, and $\sigma_\mathcal{N}$ are signs determined by
\begin{equation}
	\sigma_\mathcal{N}=\begin{cases}
		1 & \text{$\mathcal{N}$ lies in the past of $M$}\\
		-1 & \text{$\mathcal{N}$ lies in the future of $M$}
	\end{cases}
\end{equation}
In this form, the expression is manifestly invariant under reparametrizations $\lambda\mapsto\lambda^\prime$.


\bibliographystyle{JHEP}
\bibliography{ref_comp_rst}
\end{document}